\documentclass[aps,prd,preprint,groupedaddress,showpacs]{revtex4}

\usepackage{epsfig}
\usepackage{graphics}
\usepackage{slashed}
\usepackage{color}
\begin{document}

\leftmargin -2cm
\def\choosen{\atopwithdelims..}

\boldmath
\title{OPTICAL PROPERTIES OF LOWEST-ENERGY CARBON ALLOTROPES FROM THE
FIRST-PRINCIPLES CALCULATIONS} \unboldmath

\author{\firstname{V.A.}
\surname{Saleev}} \email{saleev@samsu.ru}
\author{\firstname{A.V.}\surname{Shipilova}} \email{alexshipilova@samsu.ru}

\affiliation{Samara National Research University, Moscow Highway,
34, 443086, Samara, Russia}

\begin{abstract}
We study the optical properties of lowest-energy carbon allotropes
in the infrared, visible and ultra-violet ranges of light in the
general gradient approximation of the density functional theory. In
our calculations we used the all-electron approach as well as the
pseudo-potential approximation. In the infrared range, the complex
dielectric functions, infrared and Raman spectra have been
calculated using CRYSTAL14 program. The electronic properties and
energy-dependent dielectric functions in the visible and ultraviolet
ranges have been calculated using VASP program. We have described
with a good accuracy experimentally known optical properties of
cubic diamond crystal. Using obtained set of relevant parameters for
calculations, we have predicted optical constants, dielectric
functions and Raman spectra for the lowest-energy hypothetical
carbon allotropes and lonsdaleite.


\end{abstract}

\maketitle

\section{Introduction}
\label{intro}
 Diamond and different forms of carbon materials
are the subject of the intensive theoretical and experimental study
[1,2]. It is well known that carbon can form $sp^2$ and $sp^3$
hybridized bonds which are realized under ambient conditions in
cubic diamond, graphite, fullerene and graphene [3]. At high
temperature and pressure graphite can be converted to cubic diamond
or to the lonsdaleite (2H hexagonal diamond) [4,5].  For a long
time, hexagonal diamond has been produced artificially by static and
shock wave compression of well-crystallized graphites [6]. Recently
it was shown that hexagonal diamond can be also obtained from cubic
diamond [7]. It was found that graphite cold-compression leads to a
creation of the new $sp^3$-bonded stable forms of carbon allotropes
[8]. Nowadays, there are about two hundreds of different predicted
hypothetical $sp^3$ carbon allotropes, which are collected in the
SACADA database [9]. The small difference of energy of several these
allotropes relatively to the diamond, $0.01-0.10$ eV per atom,
raises the idea of a possibility to find these allotropes in the
mixed carbon phases. The experimental search of such carbon
allotropes should be based on some new physical information about
formation of these new phases. The most probable signals can be
connected with optical properties of the materials, such as Raman
spectra, optical coefficients in the regions of infrared (IR),
visible and ultraviolet (UV) spectra, as well as  the energy
dependence of the absorption and refractive indices. Here we
calculate different optical properties of the six lowest-energy
$sp^3$ carbon allotropes: cubic diamond, lonsdaleite [4], 4H-diamond
[10], SiC12 [11], C28 [12] and mtn [13]. The main goal of our study
is to find the quantitative level of difference between optical
properties of cubic diamond and relevant carbon allotropes which can
be measured experimentally. We start with a prediction of Raman
shift spectra and IR active mode spectra, than we will study
energy-dependent complex dielectric functions and their derivative
in the ranges of visible and ultraviolet  light. Our calculations
are based on density-functional-theory (DFT) methods [14,15] as it
is implemented in CRYSTAL14 [16] and VASP [17] program packages. The
calculations for the IR range have been done using CRYSTAL14
program, which uses the all-electron approach with atomic orbital
basis sets.  The relevant calculations in the visible and
ultraviolet ranges of light have been done using VASP program
package which uses plane-wave basis sets and the pseudo-potential
approach.

\section{Computational Methods and Details}
\label{methods} Raman and IR spectra of crystal structure are
defined by the set of harmonic phonon frequencies at the $\Gamma$
point which can be obtained from the diagonalization of the Hessian
matrix of the second derivatives with respect to atomic
displacements [18]:
\begin{equation}
H^{\Gamma}_{ai,bj}=\frac{1}{\sqrt{M_a M_b}}\left( \frac{\partial^2
E}{\partial u_{ai}\partial u_{bj}} \right)
\end{equation}
where $u_{ai}$ and $u_{bj}$ are displacements of atoms $a$ and $b$
in the reference cell  along the $i$-th and $j$-th Cartesian
directions, respectively. The Raman intensity of the Stokes line of
a phonon mode $Q_p$, active due to the $\alpha_{ij}$ component of
the polarizability tensor $\alpha$, is expressed as follows:
\begin{equation}
I^p_{ij} \propto \left( \frac{\partial\alpha_{ij}}{\partial
Q_p}\right)^2
\end{equation}
The scheme of calculation, recently implemented in the CRYSTAL14
program [16], explores second-order Coupled-Perturbed
Hartree-Fock/Kohn-Sham (CPHF/KS) equations [19]. The Raman spectrum
is then computed by considering the transverse optical (TO) modes
and by adopting a pseudo-Voigt functional form: a linear combination
of a Lorentzian and a Gaussian curve with full width at half maximum
of 8 cm$^{-1}$. Raman intensities are normalized so that the largest
value is conventionally set to 100 a.u. To calculate IR spectra we
should know the complex dielectric tensor $\varepsilon_{ii}(\nu)$
which is computed for each inequivalent polarization direction on
the basis of a classical Drude-Lorentz model:
\begin{equation}
\varepsilon_{ii}(\nu)=\varepsilon_{opt,ii}+\sum_p\frac{f_{p,ii}\nu^2_p}{\nu_p^2-\nu^2-i\nu
\gamma_p}
\end{equation}
where $ii$ indicates the polarization direction,
$\varepsilon_{opt,ii}$ is the optical dielectric tensor, $\nu_p$,
$f_p$ and $\gamma_p$ are the TO frequency, oscillator strength and
damping factor for the $p$-th vibration mode, respectively. The real
and imaginary parts of $\varepsilon_{ii}(\nu)$ are computed and the
maxima of this function correspond to the TO frequencies. The
optical or high-frequency dielectric tensor is computed in a
quasi-free electron approximation via coupled perturbed Hartree-Fock
(Kone-Sham) method [20]. The refractive $(n)$ and absorption $(k)$
indices are computed as real and imaginary parts of the complex
refractive index $n^*(\nu)=\sqrt{\varepsilon(\nu)}$, also for each
inequivalent polarization direction.

In the projector augmented plane wave method realized in VASP the
frequency-dependent dielectric functions is obtained in the random
phase approximation, where the imaginary part of the
frequency-dependent dielectric tensor is written as
\begin{equation}
\varepsilon^{(2)}_{ij}(\omega)=\frac{4\pi^2 e^2}{\Omega}\lim_{q\to
0}\frac{1}{q^2}\sum_{c,v,k} 2w_k \delta(\epsilon_{ck}-\epsilon_{v
k}-\omega)\langle u_{ck+e_i q}| u_{v k}\rangle \langle u_{ck+e_j
q}|u_{v k} \rangle ^*
\end{equation}
where the indices $c$ and $v$ refer to conduction and valence band
states respectively in the sum over the empty states,
$\epsilon_{c,vk}$ are the corresponding eigenenergies, $\Omega$ is
the volume of a primitive cell, $k-$point weights $w_k$ are defined
such that they sum to 1, $e_{i,j}$ are the unit vectors for the
three Cartesian directions,  and  $u_{ck}$ is the cell-periodic part
of the orbitals at point $k$. The real part of the dielectric tensor
$\varepsilon^{(1)}(\omega)$ is obtained by the Kramers-Kronig
transformation
\begin{equation}
\varepsilon_{ij}^{(1)}=1+\frac{2}{\pi}P \int_0^\infty
\frac{\varepsilon^{(2)}_{ij}(\omega')\omega'
d\omega'}{\omega'^2-\omega^2+i\eta}
\end{equation}
where $P$ denotes the principal value. By cubic symmetry, the
following relation is satisfied for diamond and $mtn$ allotrope
\begin{equation}
\varepsilon_{xx}^{(1,2)}=\varepsilon_{yy}^{(1,2)}=\varepsilon_{zz}^{(1,2)},
\quad \varepsilon_{ij}^{(1,2)}=0, i\neq j,
\end{equation}
so the real and imaginary parts of the complex dielectric function
$\varepsilon=\varepsilon_1+i\varepsilon_2$ can be determined by
$\varepsilon_{1,2}=\varepsilon_{xx}^{(1,2)}$. In the case of the
other, anisotropic, structures we use the average values:
\begin{equation}
\varepsilon_{1,2}=\overline{\varepsilon_{1,2}}=\frac{1}{3}\left(
\varepsilon_{xx}^{(1,2)}+ \varepsilon_{yy}^{(1,2)}+
\varepsilon_{zz}^{(1,2)}  \right).
\end{equation}

\section{Computational setup}
\label{setup}
\subsection{CRYSTAL14}

We use the two schemes of DFT calculations, both are implemented in
the CRYSTAL14 program [16]. The first one incorporates the
Perdew-Burke-Ernzerhof (PBE) exchange-correlation functional [21] at
general gradient approximation (GGA) of DFT and POB-TZVP
all-electron basis set [22], and the second one – the B3LYP hybrid
functional [23] and adopted Pople's 6-21G Gaussian all-electron
basis set [24,25].   The level of accuracy of calculating the
energies of Coulomb and Hartree-Fock exchanges is controlled by a
set of TOLINTEG parameters, which were chosen as \{8, 8, 8, 8, 18\}.
The convergence threshold on energy for the self-consistent-field
(SCF) calculations is $10^{-7}$ Hartree for structural optimization
and $10^{-8}$ Hartree for vibration frequency calculations.  The
number of basis vectors in the irreducible Brillouin zone is given
by the shrink parameter IS = 8 for structural optimization and IS=16
for vibration frequency calculations. The relaxation of cell
parameters and atomic positions to equilibrium values was carried
out until the lattice stress became less than 0.02 GPa.

\subsection{VASP}

The computations of the complex dielectric functions  in the visible
and ultraviolet regions were performed by the VASP package [17] at
the level of GGA and with two functionals: the above-mentioned PBE
[21] and the screened Heyd, Scuseria, and Ernzerhof (HSE06) hybrid
functional [26], since the latter is well-suited to reproduce the
electronic properties including the band gaps of a number of
elemental and binary insulators and semiconductors. For this
purpose, at first we performed the relaxation of the discussed
structures with a cutoff energy of 700 eV for the plane-wave basis
set, until the energy was converged up $10^{-7}$ eV per unit cell
and the residual stress was less than $10^{-5}$ eV per atom.  The
integration over the Brillouin zone was performed using
Monkhorst-Pack grids with the number of k-points along each
direction of the reciprocal cell equal to an integer divisor of 25
{\AA} over the length of corresponding lattice vector. The chosen
set of input parameters gives the reliable results for the main
physical properties of diamond as it was shown in our previous work
[12].

\section{Structural properties}
\label{structure} 
\begin{figure}[p]
\begin{center}
\includegraphics[width=.43\textwidth, angle=270, clip=]{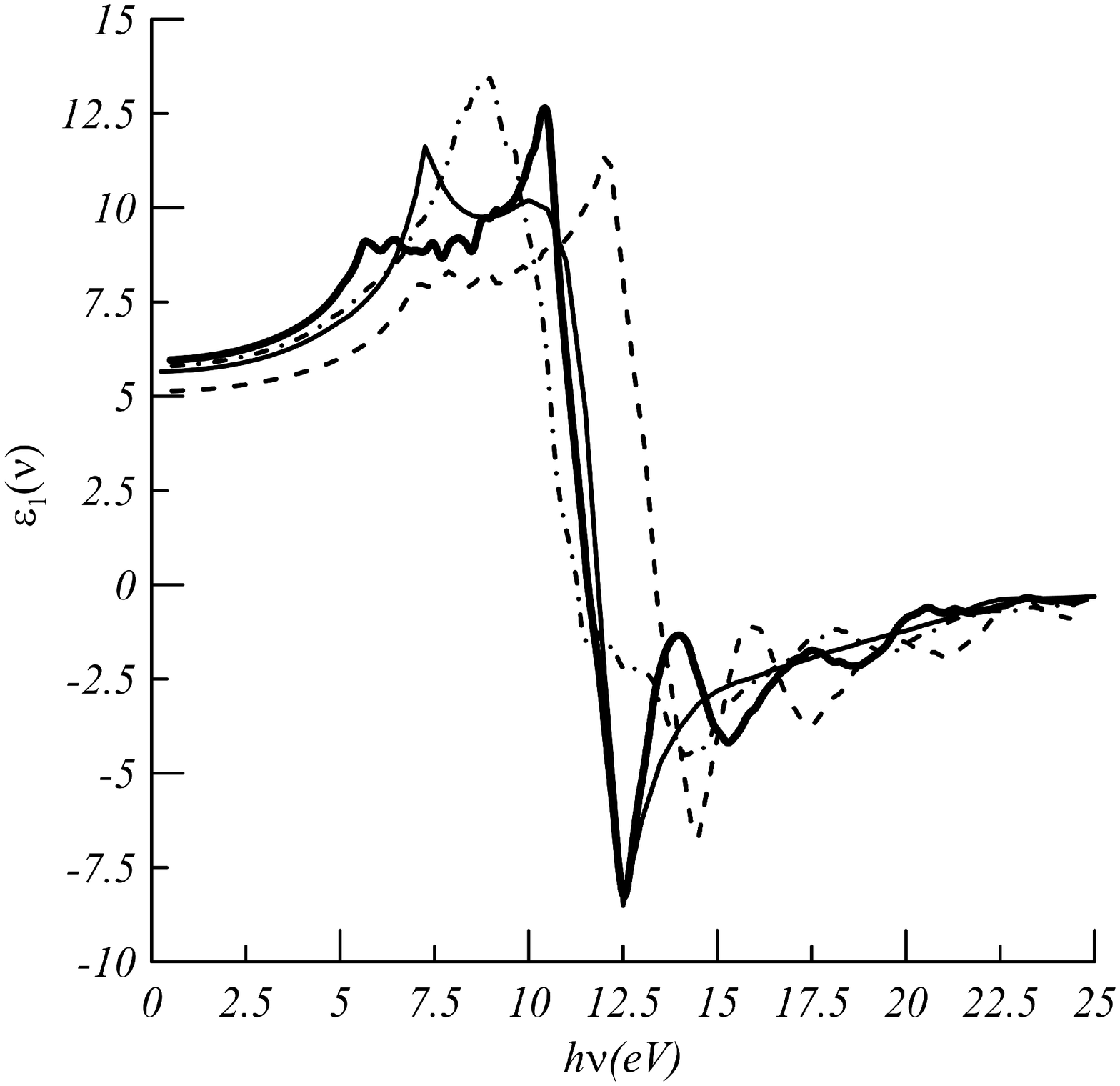}
\includegraphics[width=.43\textwidth, angle=270, clip=]{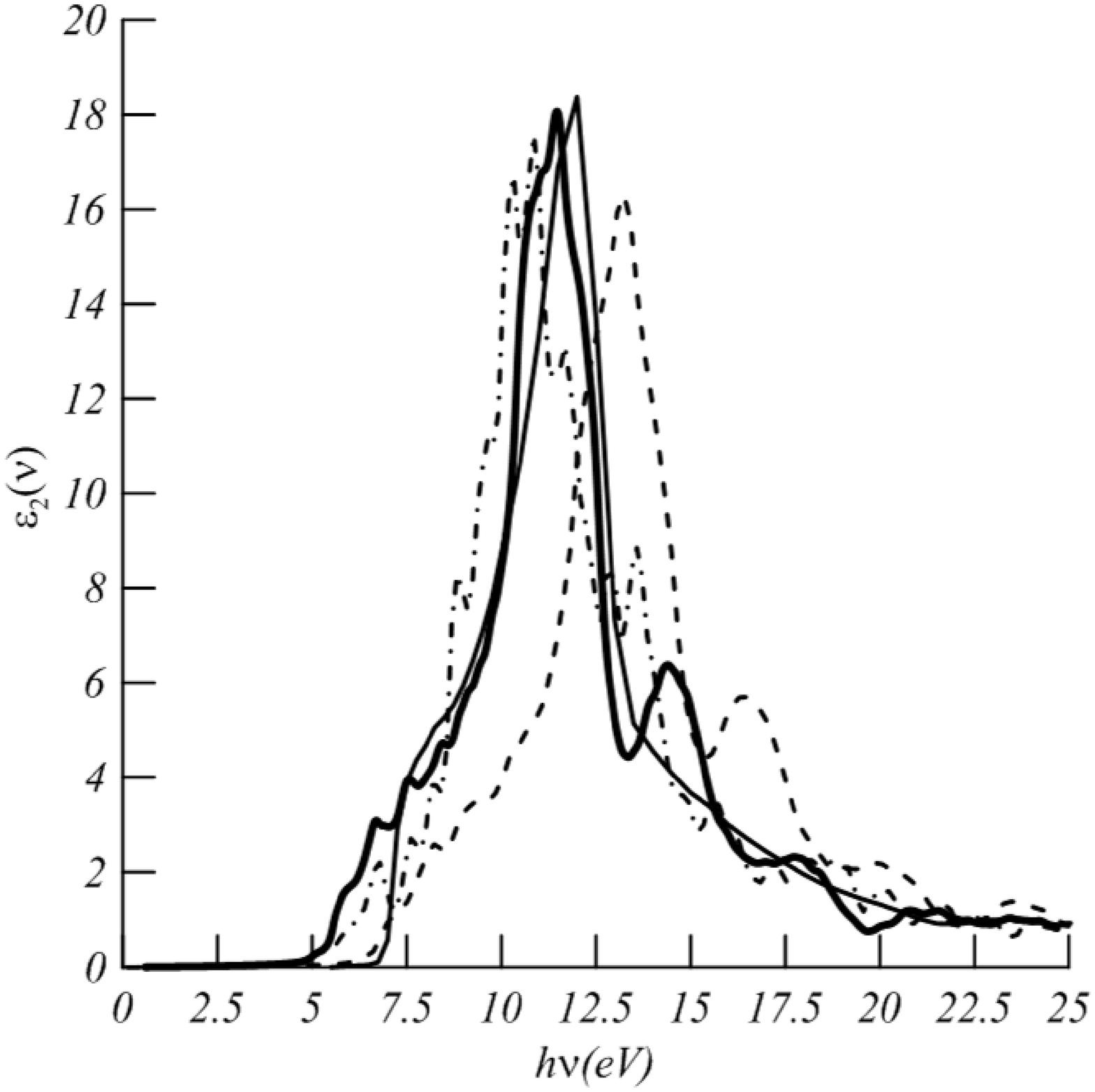}
\caption{Real (a) and imaginary (b) dielectric functions: solid line
- experiment for cubic diamond, bold solid line - prediction for
cubic diamond with PBE exchange-correlation functional, dashed line
- prediction for cubic diamond with HSE06 functional, dash-dotted
line - lonsdaleite with PBE functional. \label{fig:4}}
\end{center}
\end{figure}

The full geometry optimization of studied
structures have been performed using the quasi-Newton algorithm in
CRYSTAL14 program. The relaxed lattice constants and atomic Wyckoff
positions are presented in the Tables \ref{Table:6} and
\ref{Table:7}. The differences between lattice constants obtained in
two schemes of calculations are about 1 \%. While the energy
differences per atom relatively diamond are equal, see Tables
\ref{Table:1} and \ref{Table:2}. To check a mechanical stability of
hypothetical allotropes under study, we calculated the elastic
constants and phonon spectra which are collected in the
Fig.~\ref{fig:4}. Our results for structural properties agree with
previous DFT calculations [10-13].

\section{Raman and IR spectra}
\label{raman} 
\begin{figure}[p]
\begin{center}
\includegraphics[width=.45\textwidth, clip=]{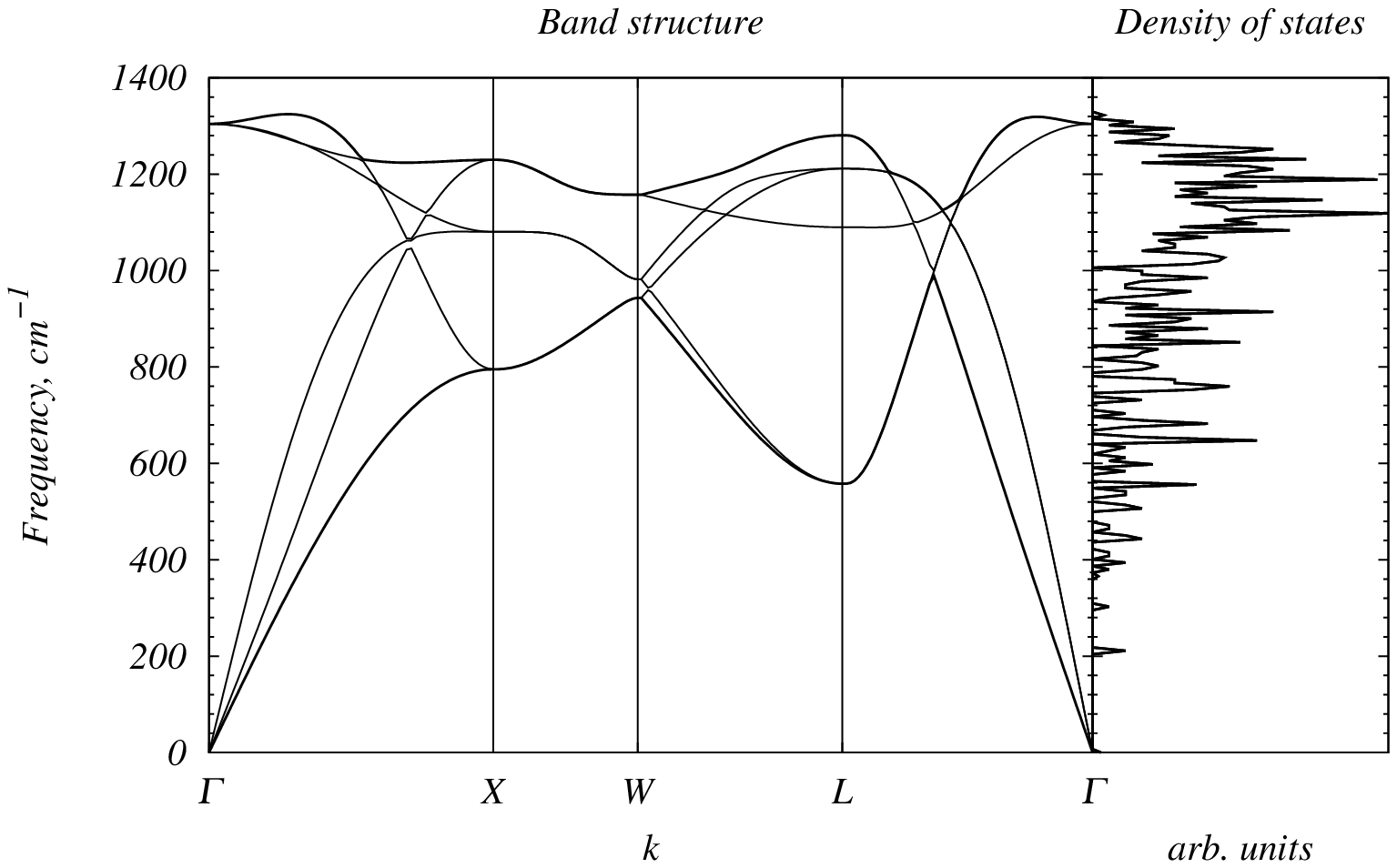}
\includegraphics[width=.45\textwidth, clip=]{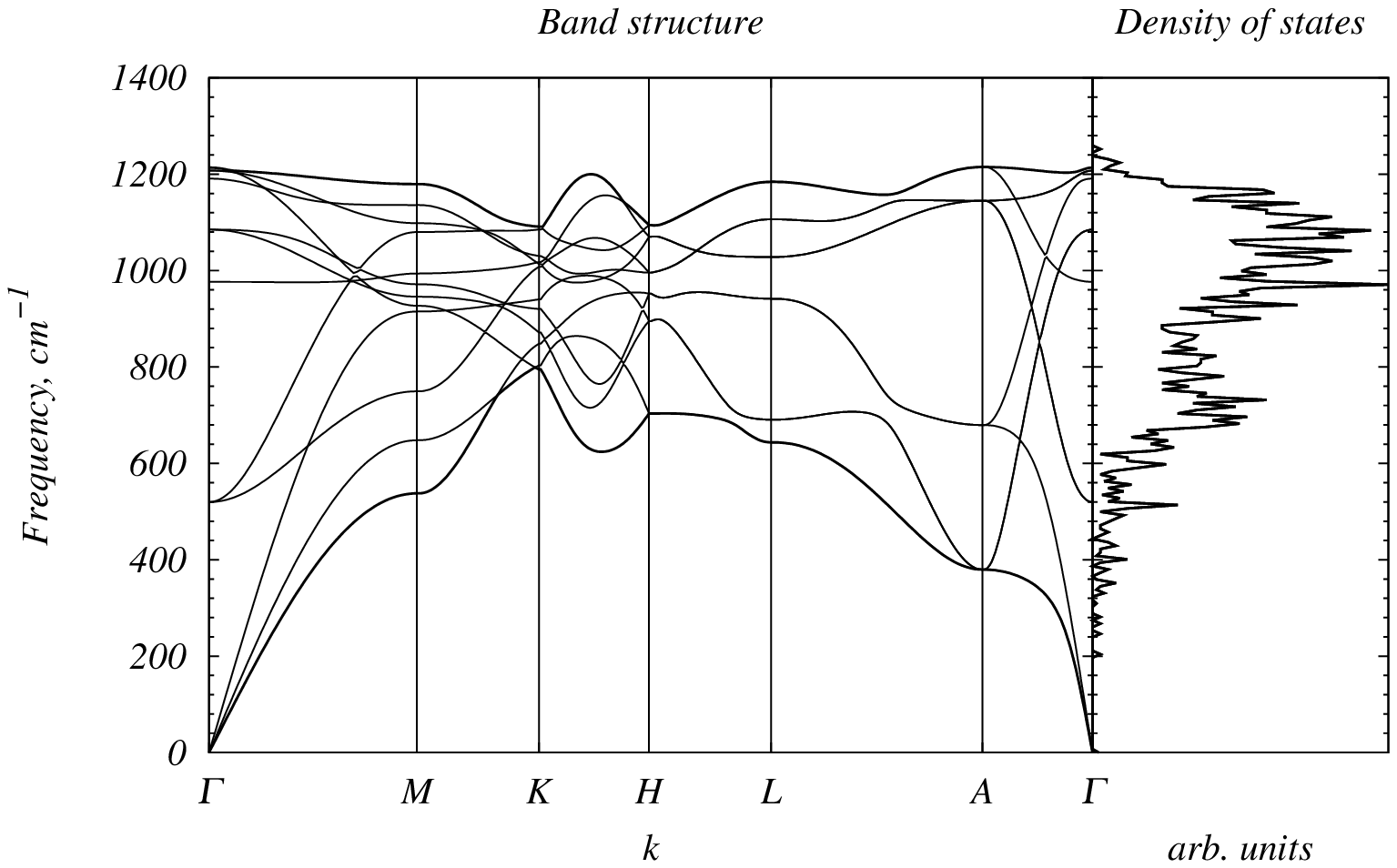}
\includegraphics[width=.45\textwidth, clip=]{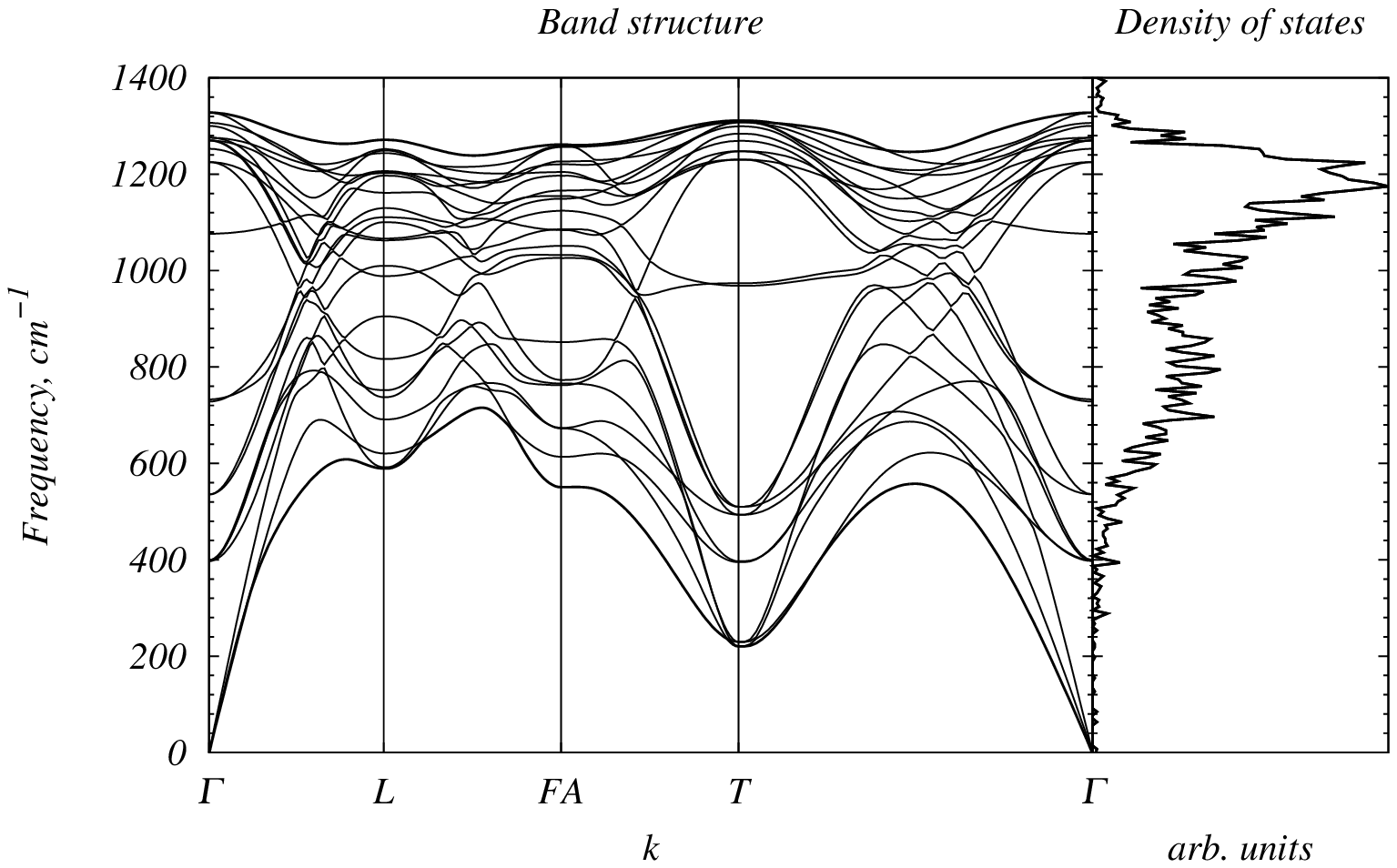}
\includegraphics[width=.45\textwidth, clip=]{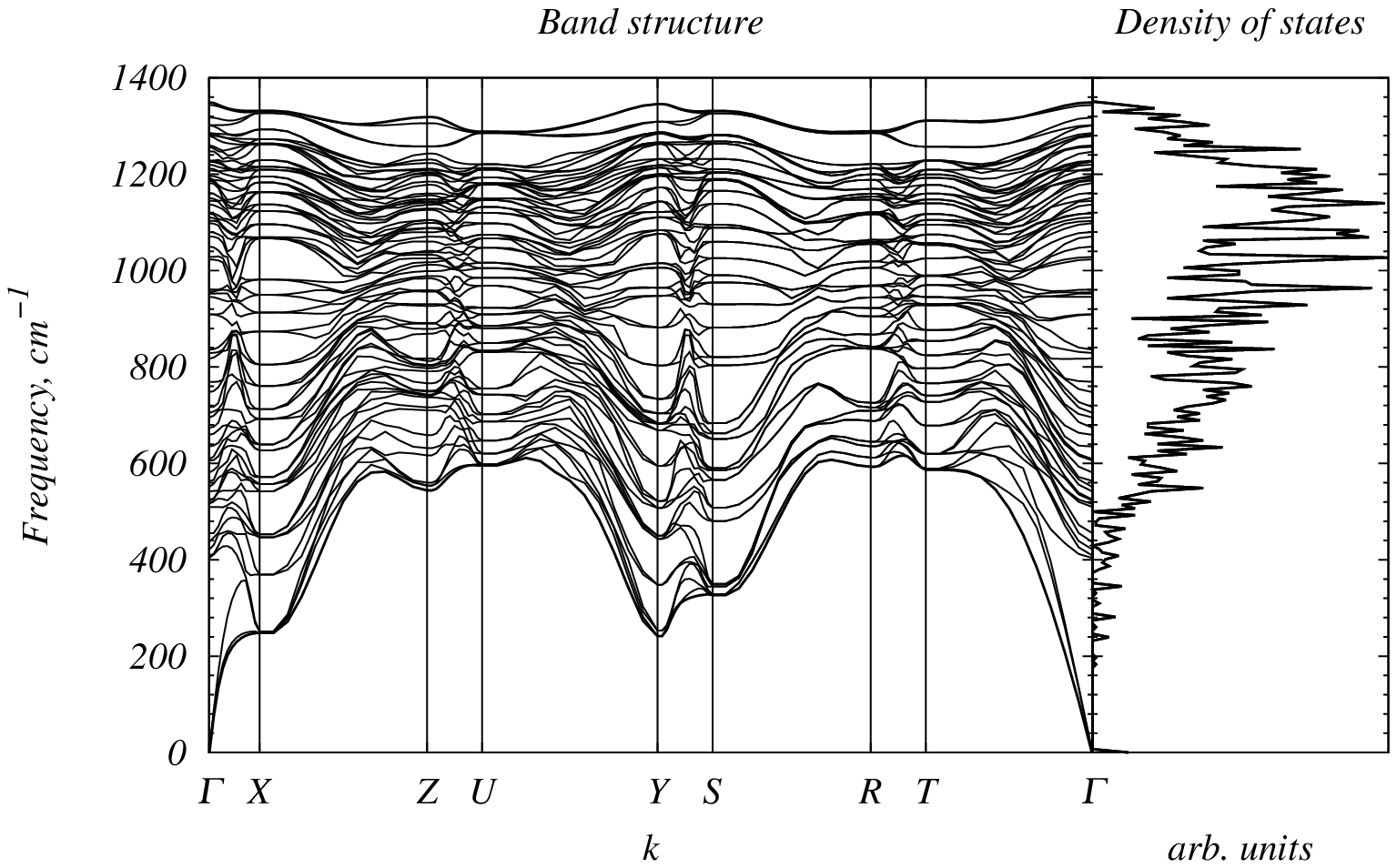}
\includegraphics[width=.45\textwidth, clip=]{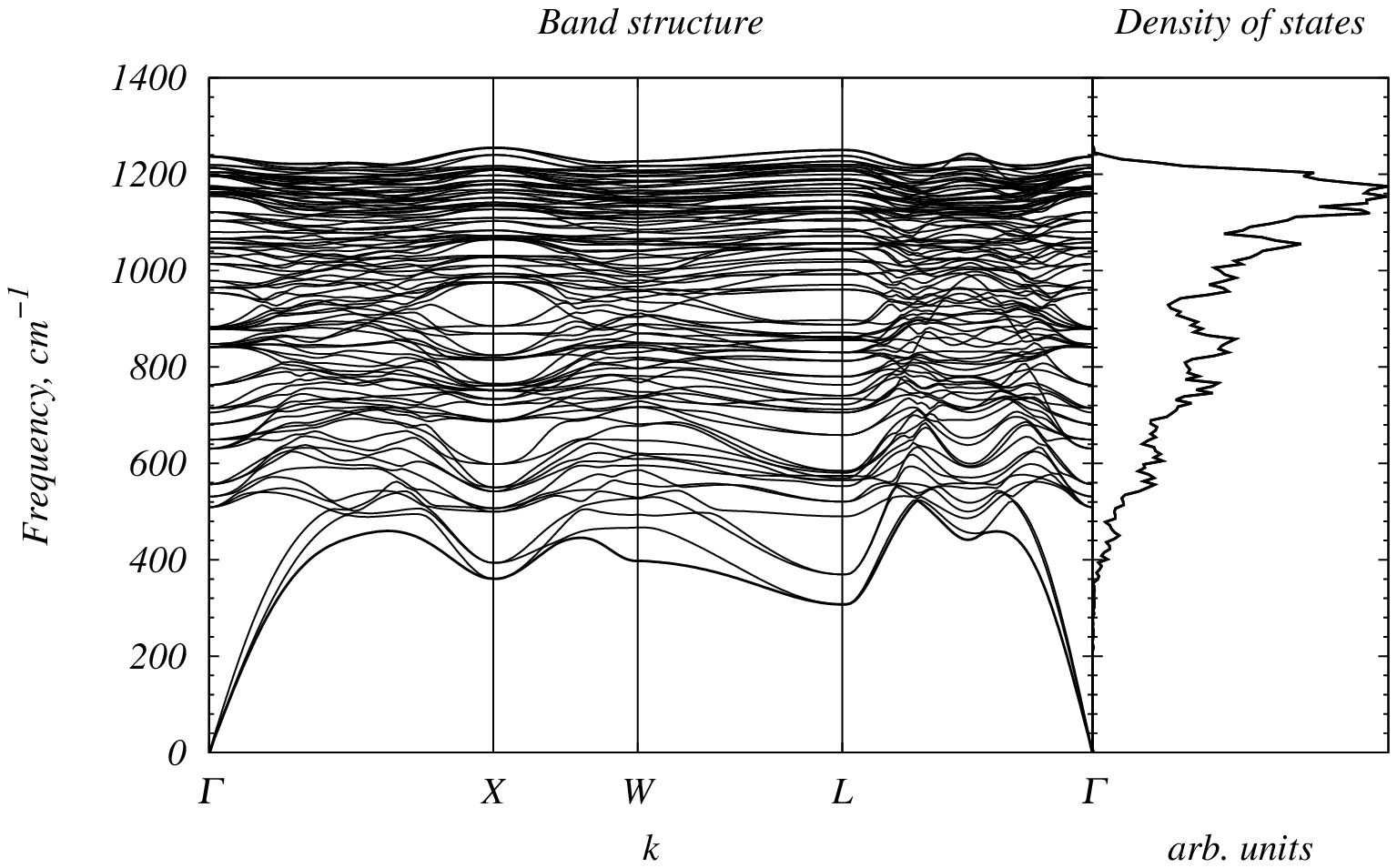}
\includegraphics[width=.45\textwidth, clip=]{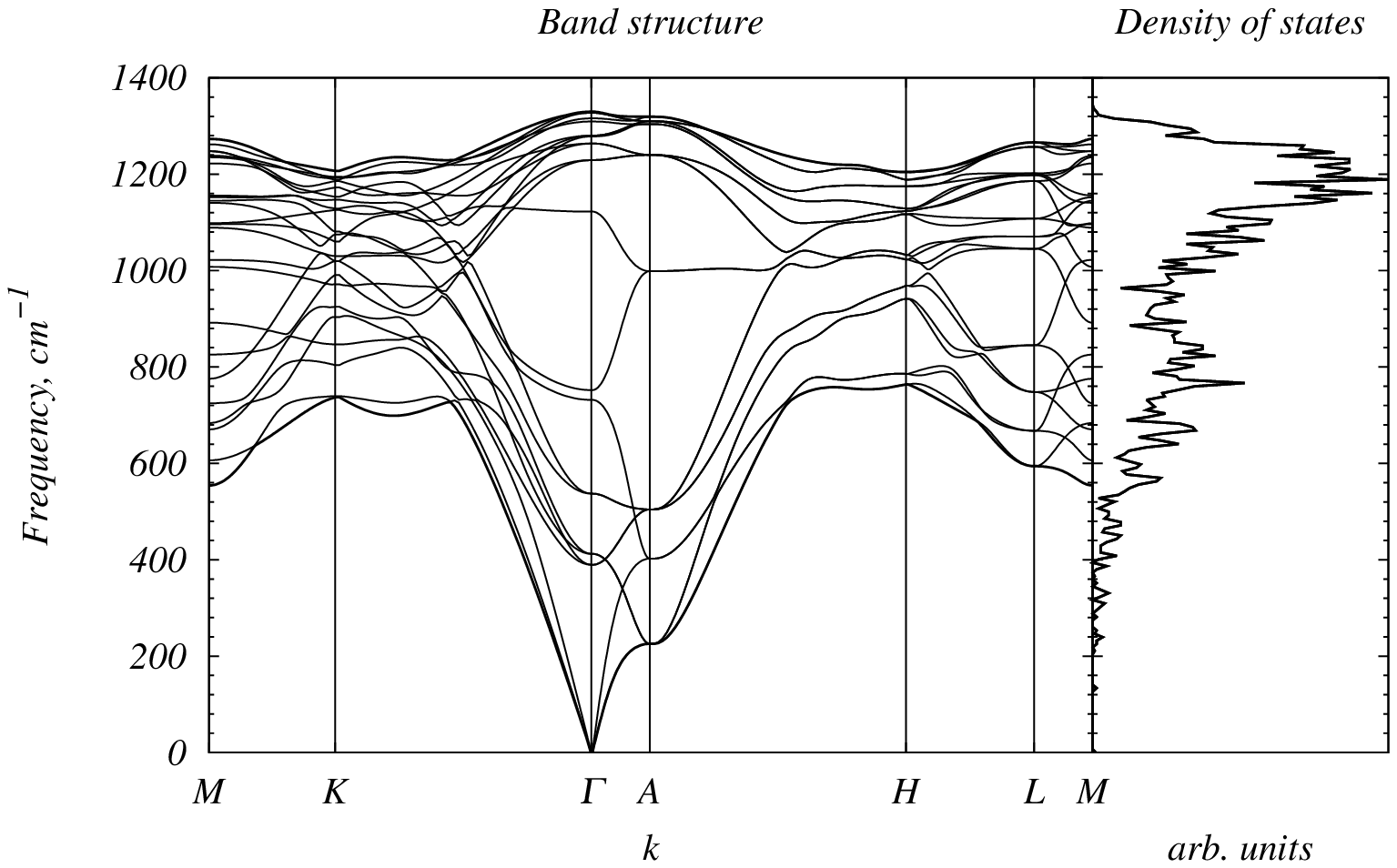}
\caption{Phonon band structures and density of states of
  allotropes. Left column: diamond, SiC12, and mtn. Right column:
lonsdaleite, C28, and 4H-diamond}\label{fig:1}
\end{center}
\end{figure}

Raman identification of new carbon allotropes needs a
precise calculation of vibration spectra. We performed these
calculations applying quasi-harmonic approximation at the point as
it was done for Raman identification of lonsdaleite in Refs. [27,28]
and for Raman spectroscopy of nanocrystalline diamond in Ref. [29].
We have found for cubic diamond that the position of single Raman
peak, corresponding to the first-order scattering of $F_{2g}$
symmetry, is 1297 cm$^{-1}$ (PBE, POB-TZVP ) and 1332 cm$^{-1}$
(B3LYP, 6-21G). The last one is in a best agreement with
experimental value. Such a way, we perform calculations of Raman and
IR spectra here and after using B3LYP exchange-correlation
functional and modified Pople's 6-21 basis set [25]. The presented
in Fig.~\ref{fig:1} polycrystalline (powder) Raman spectra were
computed by averaging over the all possible orientations of the
crystallites.

For the lonsdaleite we predict three Raman active modes: $E_{1g}$
(1336), $A_{1g}$ (1312) and $E_{2g}$ (1209). Up to now, the pure
monocrystalline lonsdaleite have not been found or synthesized and
there is a well-known problem of correct diagnostics of lonsdaleite
phase within cubic diamond-lonsdaleite intergrowths. Possible
attendance of nanocrystalline cubic diamond can be a reason of
incorrect interpretation of Raman spectra in experimental study. It
was found that the Raman modes dynamics under laser heating allows
splitting of cubic diamond and lonsdaleite modes and recognize
diamond polymorphs [28]. The measurements in the region of 1300-1340
cm$^{-1}$ show the presence of two Raman active modes, $A_{1g}$
(1319) and $F_{1g}$ (1322). The control measurement for co-sized
cubic diamond particles give downshift of a Raman active mode
$F_{2g}$ (1332) up to 1326 cm$^{-1}$. Taking into account this
downshift effect, we can declare the agreement of our prediction
with the experimental data for lonsdaleite [28] at least for two
high-frequency modes. We also obtain a rough agreement with recent
measurements of Ref. [29], in which the most intensive band at
1292-1303 cm$^{-1}$ and at 1219-1244 cm$^{-1}$ are interpreted as
contributions from $A_{1g}$ and $E_{2g}$   vibration modes of
lonsdaleite phase in Popigai impact rock.  Early DFT calculations,
performed in the local density approximation (LDA), gave the
following results: $E_{1g}$(1312), $A_{1g}$(1305), $E_{2g}$(1193)
[30] and $E_{1g}$(1338), $A_{1g}$ (1280), $E_{2g}$(1221) [31].

In Fig.~\ref{fig:1}(a-e), we present our results for polycrystalline (powder)
Raman spectra with intensity plotted in arbitrary units. Of course,
it has the dependence on direction for monocrystalline structures,
but it is small and we discuss it later for real refractive indices
only. The single peak of cubic diamond corresponding $F_{2g}$(1332)
mode is shown in all figures for comparison. The number of active
Raman modes or peaks depends on symmetry of crystal lattice and on
the number of non-equivalent atoms in the asymmetric unit. As we
see, quantum-mechanical approach predicts a very specific Raman
spectrum for every allotrope, which can be considered as "finger
prints" of these structures in the experimental search. The
improvement of experimental methods to the level of accuracy of
theoretical calculations will be crucial for search and recognition
of diamond polymorphs.

The high-frequency refractive indices  for different direction
(ii=xx, yy, zz), accordingly Vogt notation, are collected in the
Table~\ref{Table:3}. The calculation with PBE functional and POB-TZVP basis set
gives answer which is very close to the experimental value for cubic
diamond (~2.40) instead of the calculation with B3LYP functional
used successfully for Raman spectrum calculation. The refractive
asymmetry factor can be defined as follows. The values of  for
lonsdaleite, C28, SiC12 and 4H-diamond are about 2.5-3.5 \%.
This is enough large values for experimental test and recognition.
The list of calculated IR active modes which initiate absorption of
light in IR region are presented in Table~\ref{Table:4}. The ideal cubic diamond
and lonsdaleite crystals do not absorb IR light. It means that
experimental viewing of absorption for diamond-like structures may
be used for recognition of new $sp^3$ carbon allotropes.

\section{Visible and UV spectra}
\label{visible} 
\begin{figure}[p]
\begin{center}
\includegraphics[width=.45\textwidth, angle=270, clip=]{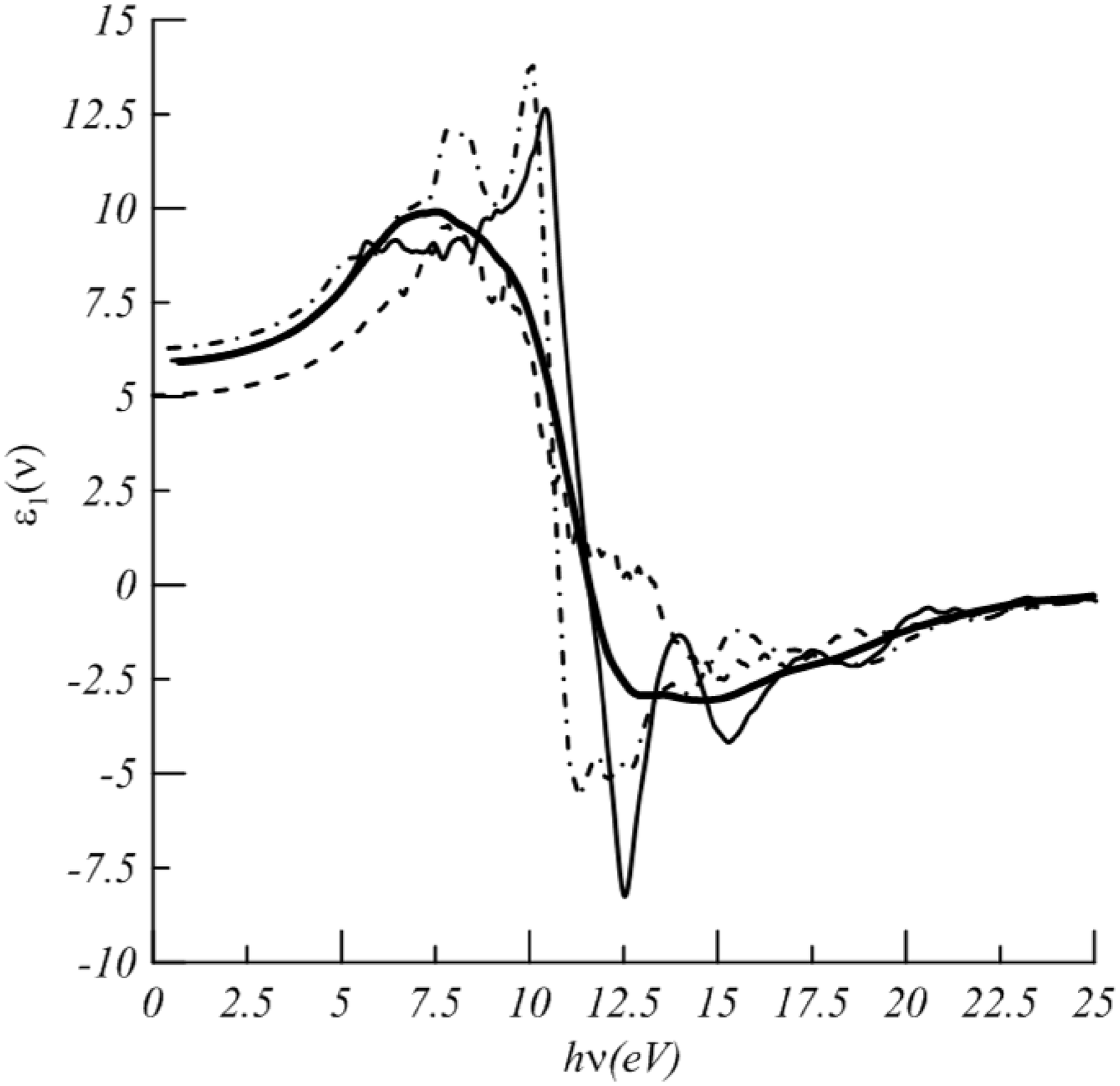}
\includegraphics[width=.45\textwidth, angle=270, clip=]{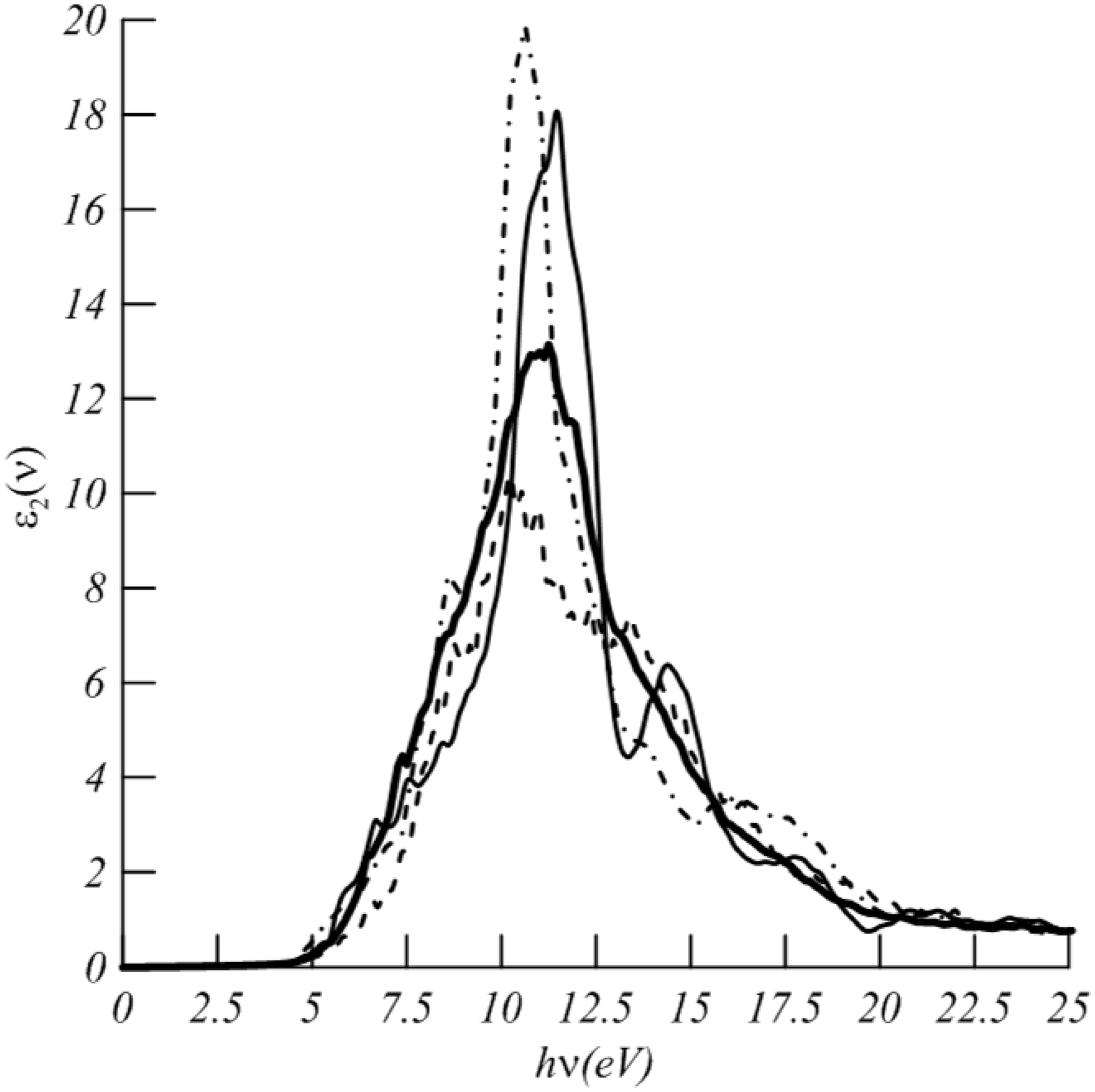}
\caption{Real (a) and imaginary (b) dielectric functions predicted
for carbon allotropes with PBE functional: solid line - cubic
diamond, bold solid line - C28, dashed line - mtn, dash-dotted line
- SiC12. \label{fig:5}}
\end{center}
\end{figure}

Optical properties are immediately connected with
the electronic band structure of the crystal. To calculate the
interband and intraband electron optical transitions one needs to
solve the band structure of the system and know its band gap in
different symmetry points of reciprocal space. Because of this
reason, we calculated the electronic band structure for all the
allotropes under study (see Fig.~\ref{fig:5}) and collected to the
table our results for indirect and direct (in $\Gamma$-point) band
gaps for all the considered structures within PBE and HSE06
functionals (see Table~\ref{Table:5}). It is well-known that the
former fails to reproduce the optical band gaps while the latter
provides the reliable results. Our result of 5.38 eV band gap for
diamond successfully reproduces the well-known experimental value of
5.47 eV [1].

\begin{figure}[p]
\begin{center}
\includegraphics[width=.4\textwidth, clip=]{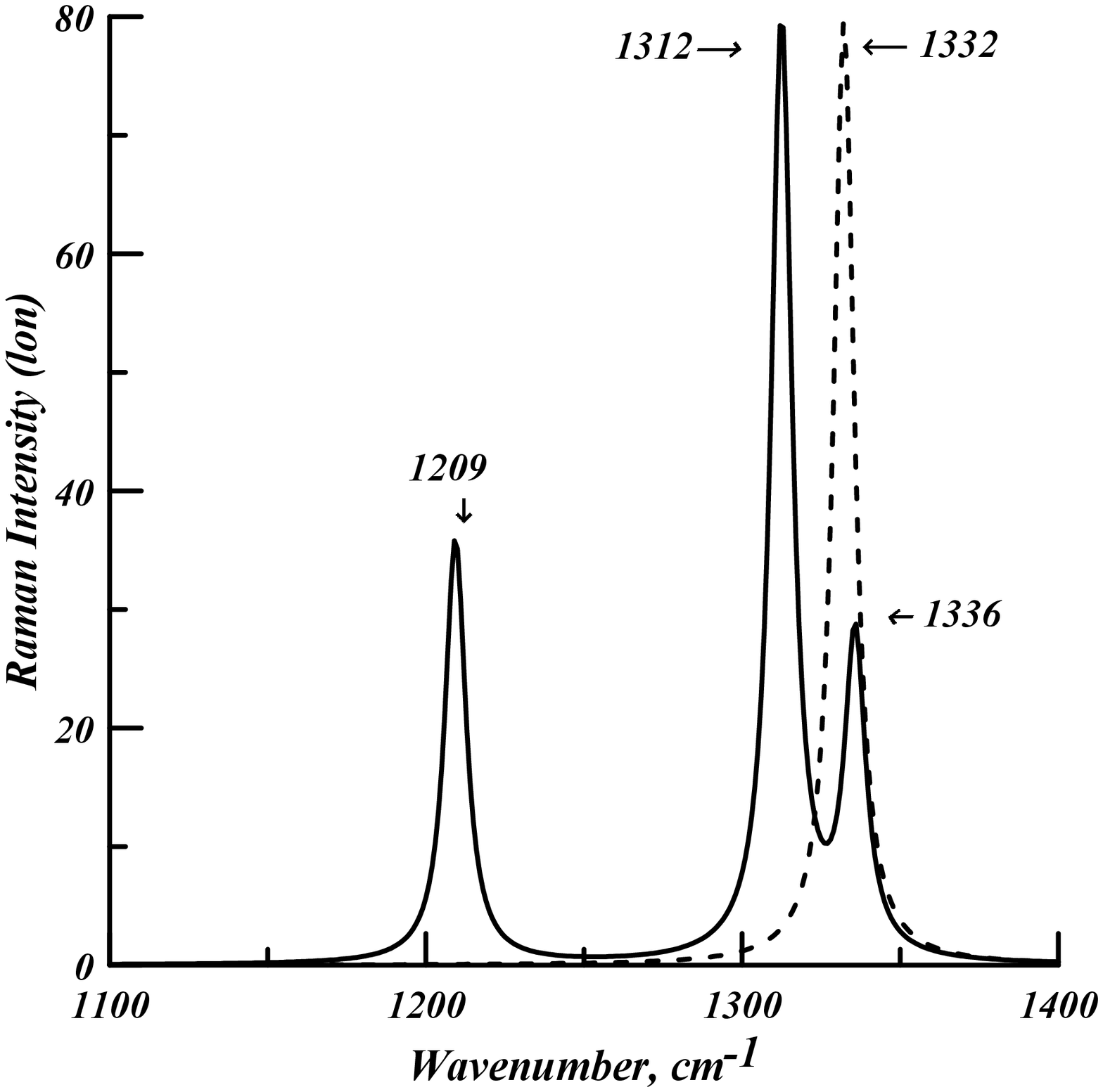}
\includegraphics[width=.4\textwidth, clip=]{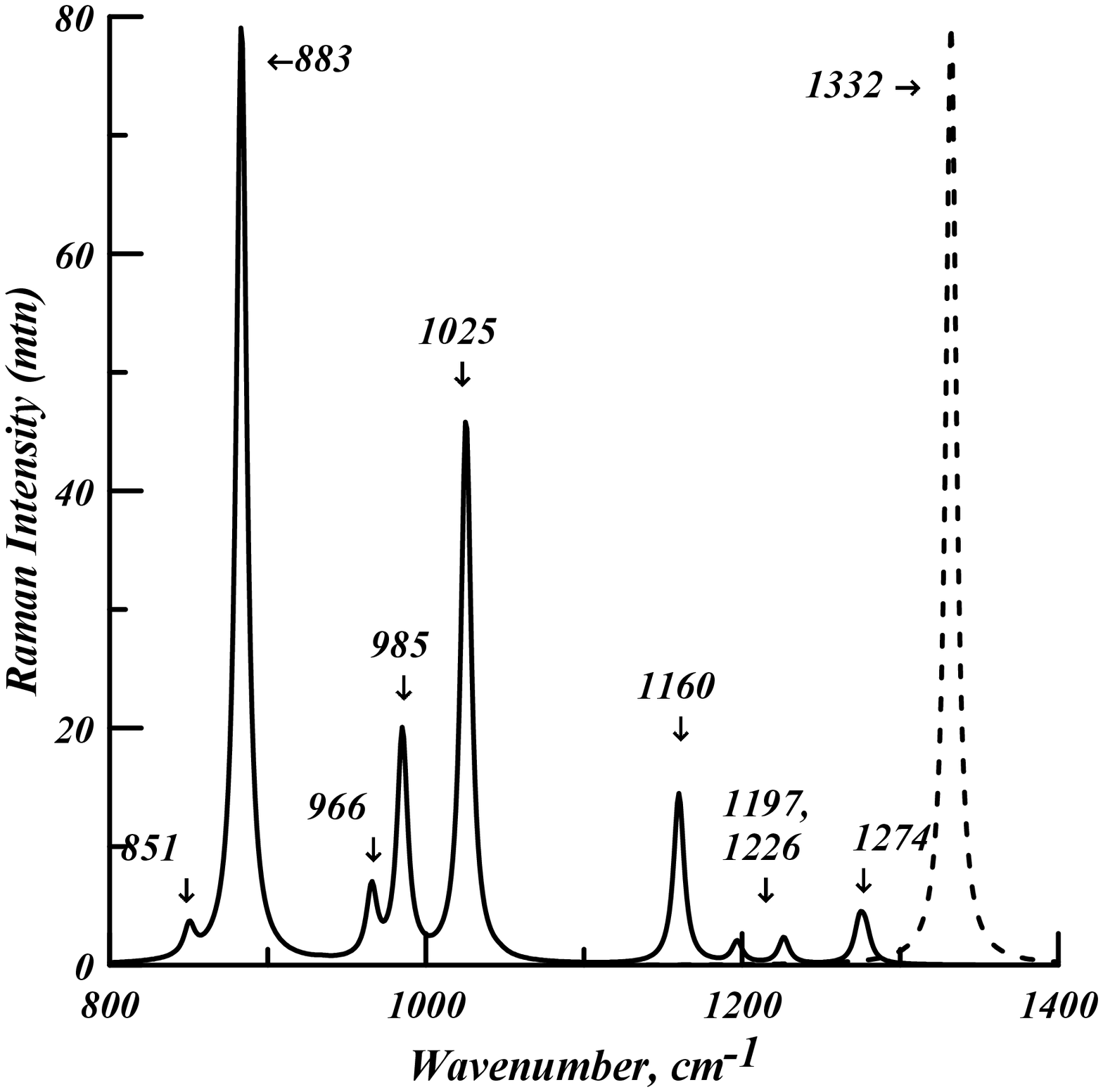}
\includegraphics[width=.4\textwidth, clip=]{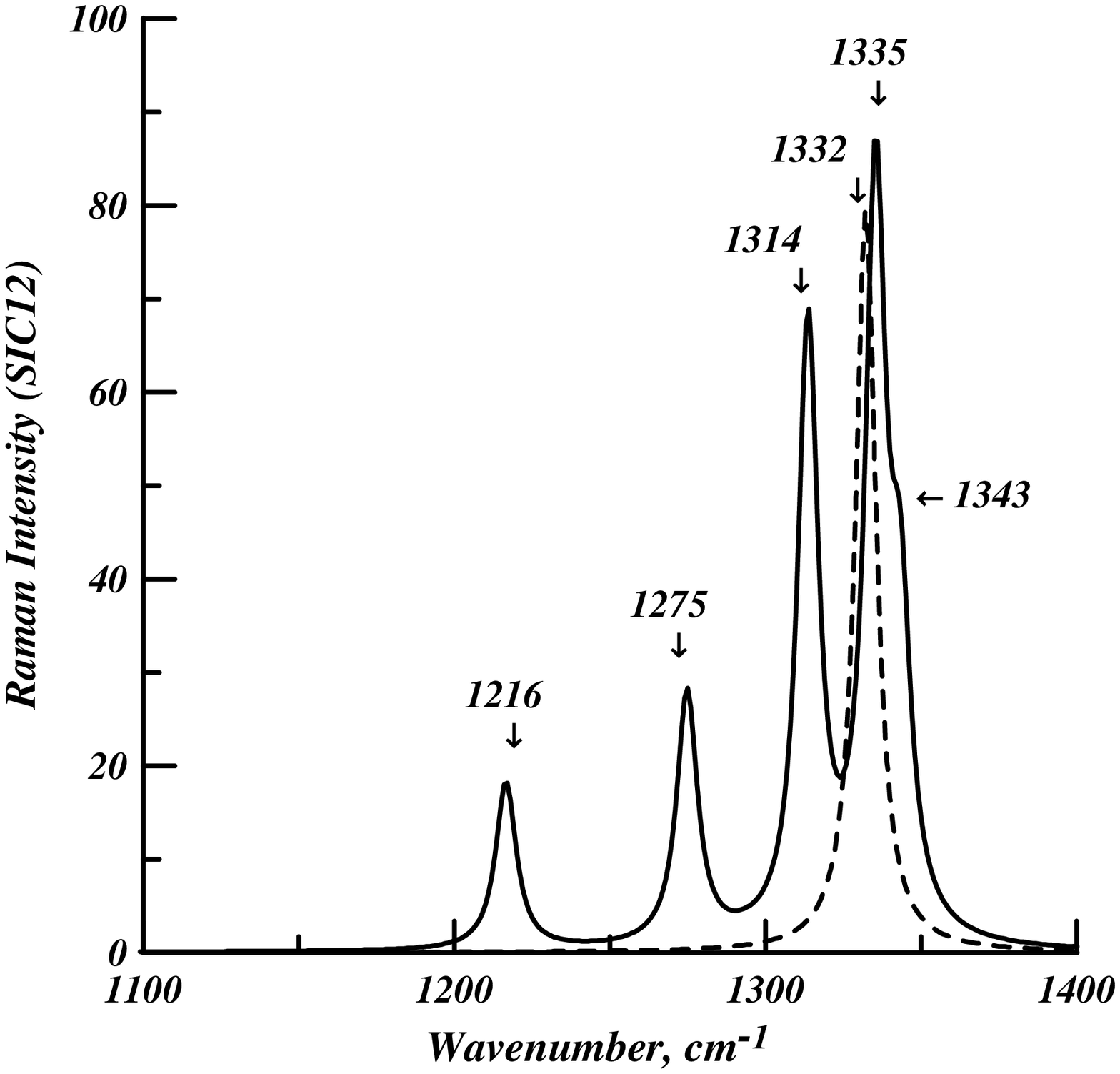}
\includegraphics[width=.4\textwidth, clip=]{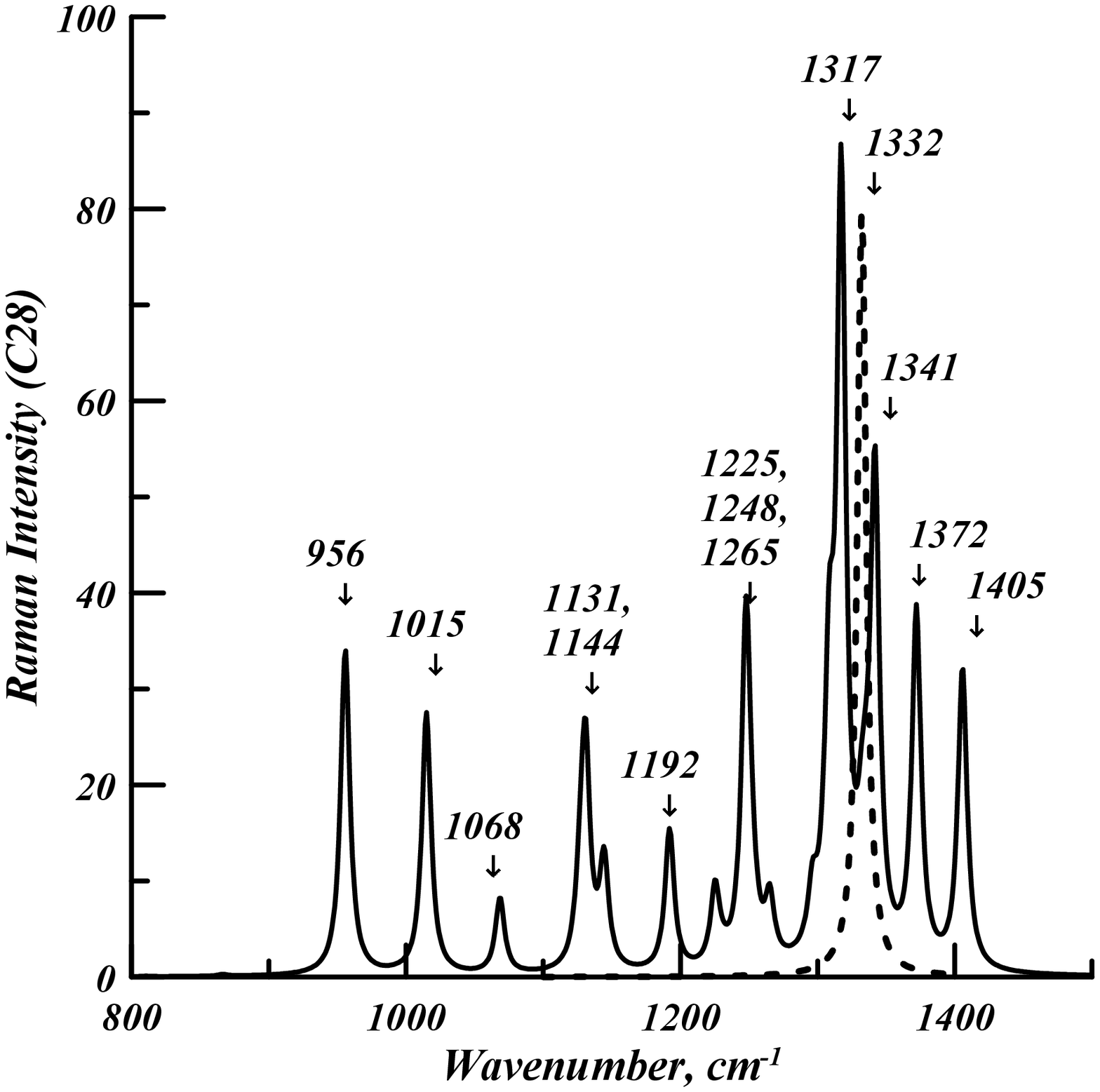}
\includegraphics[width=.4\textwidth, clip=]{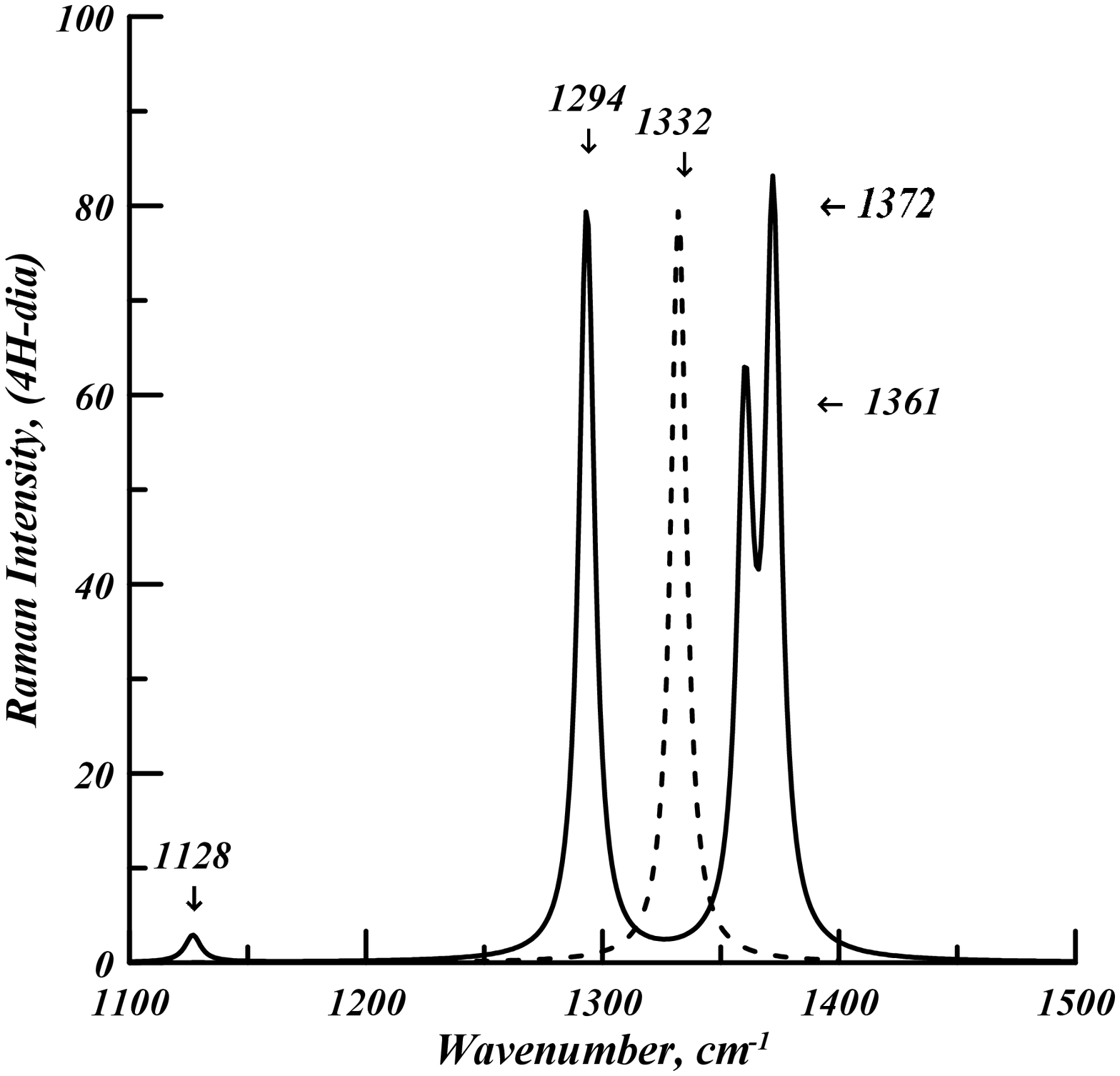}
\caption{Raman shift spectra in arbitrary units  for carbon
allotropes: a) lonsdaleite, b) mtn, c) SiC12, d) C28, e) 4H-diamond.
The peak at the 1332 cm$^{-1}$ in all panels corresponds cubic
diamond. \label{fig:2}}
\end{center}
\end{figure}

\begin{figure}[p]
\begin{center}
\includegraphics[width=.3\textwidth, angle=270, clip=]{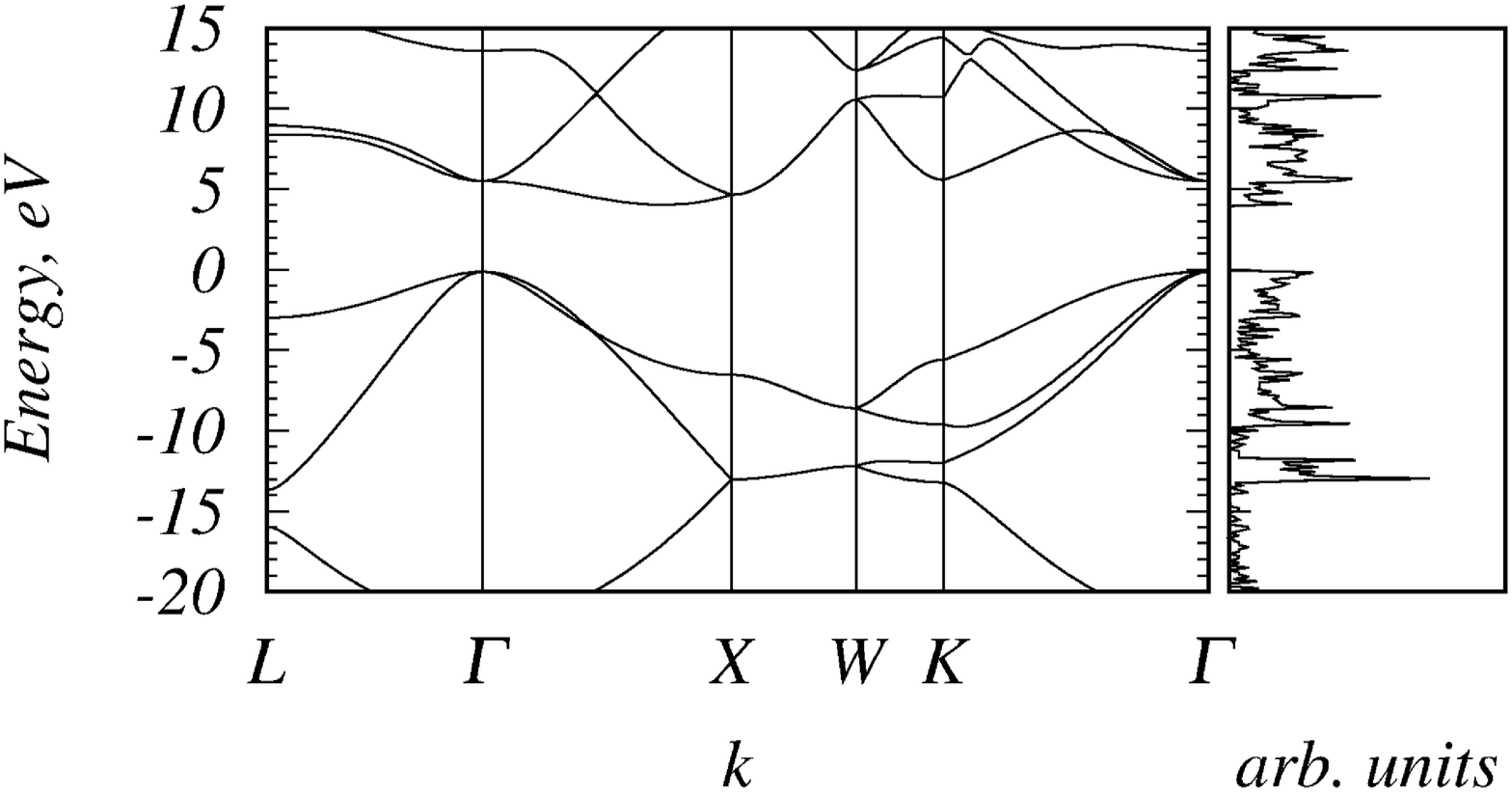}
\includegraphics[width=.3\textwidth, angle=270, clip=]{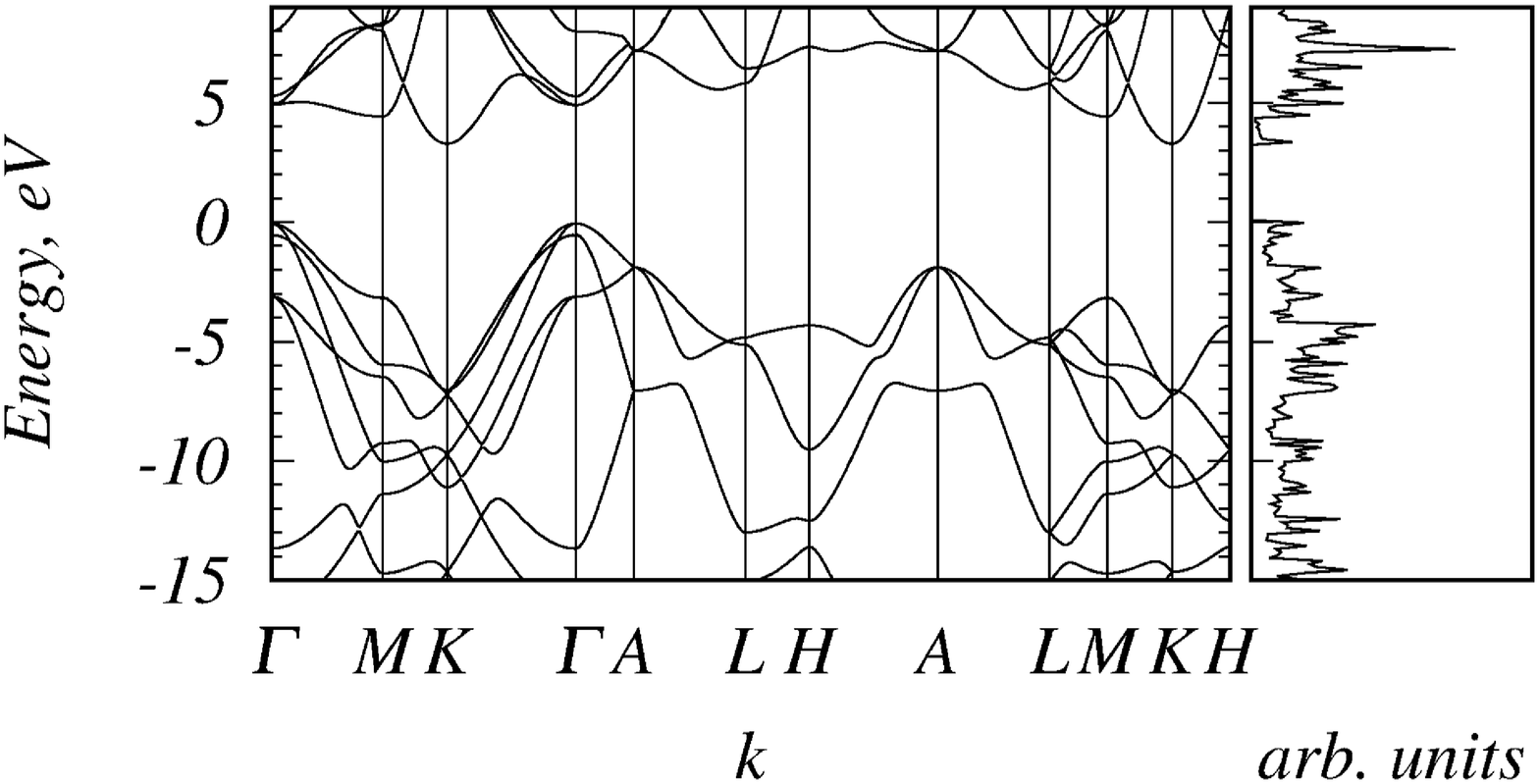}
\includegraphics[width=.3\textwidth, angle=270, clip=]{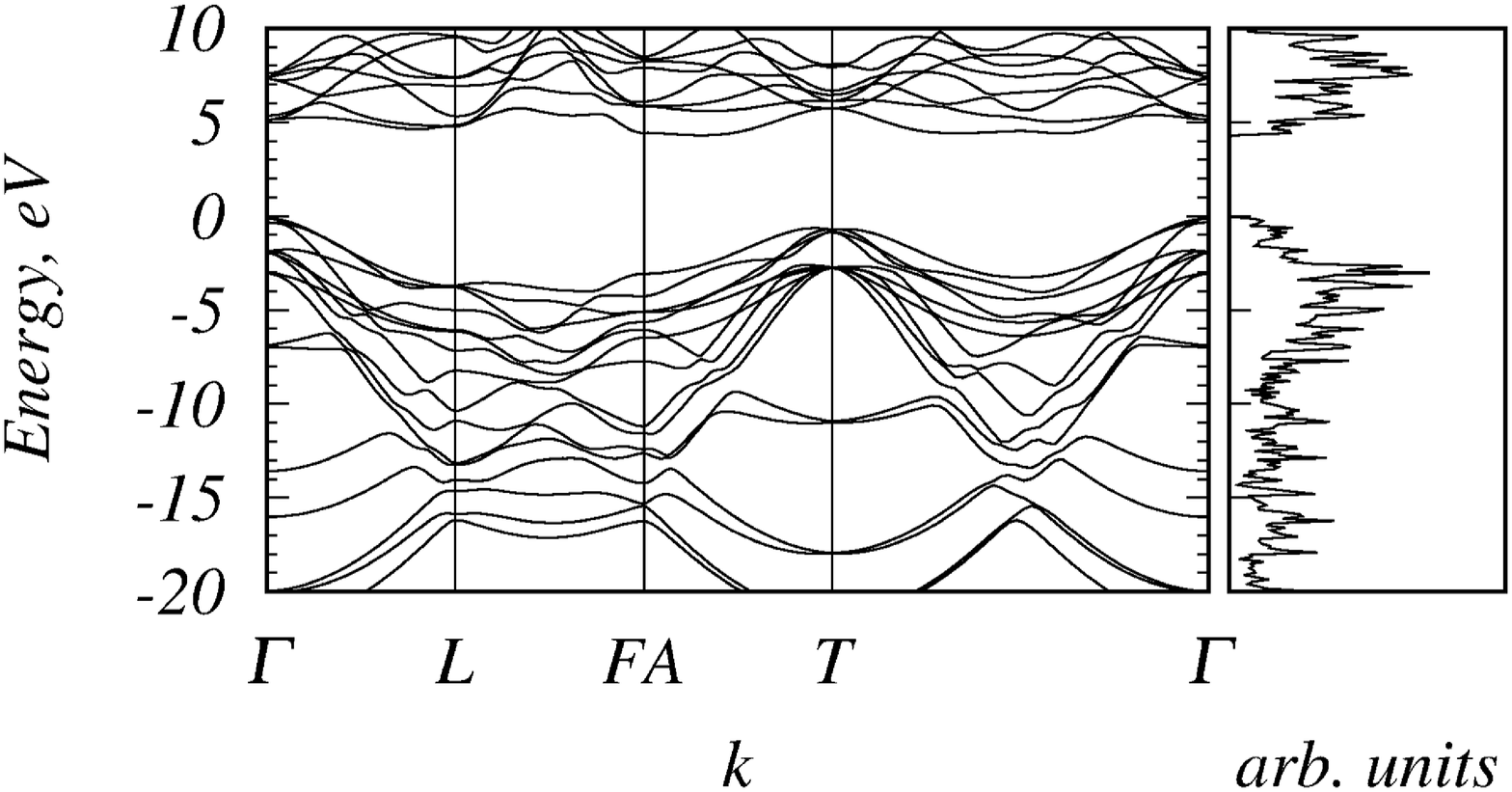}
\includegraphics[width=.3\textwidth, angle=270, clip=]{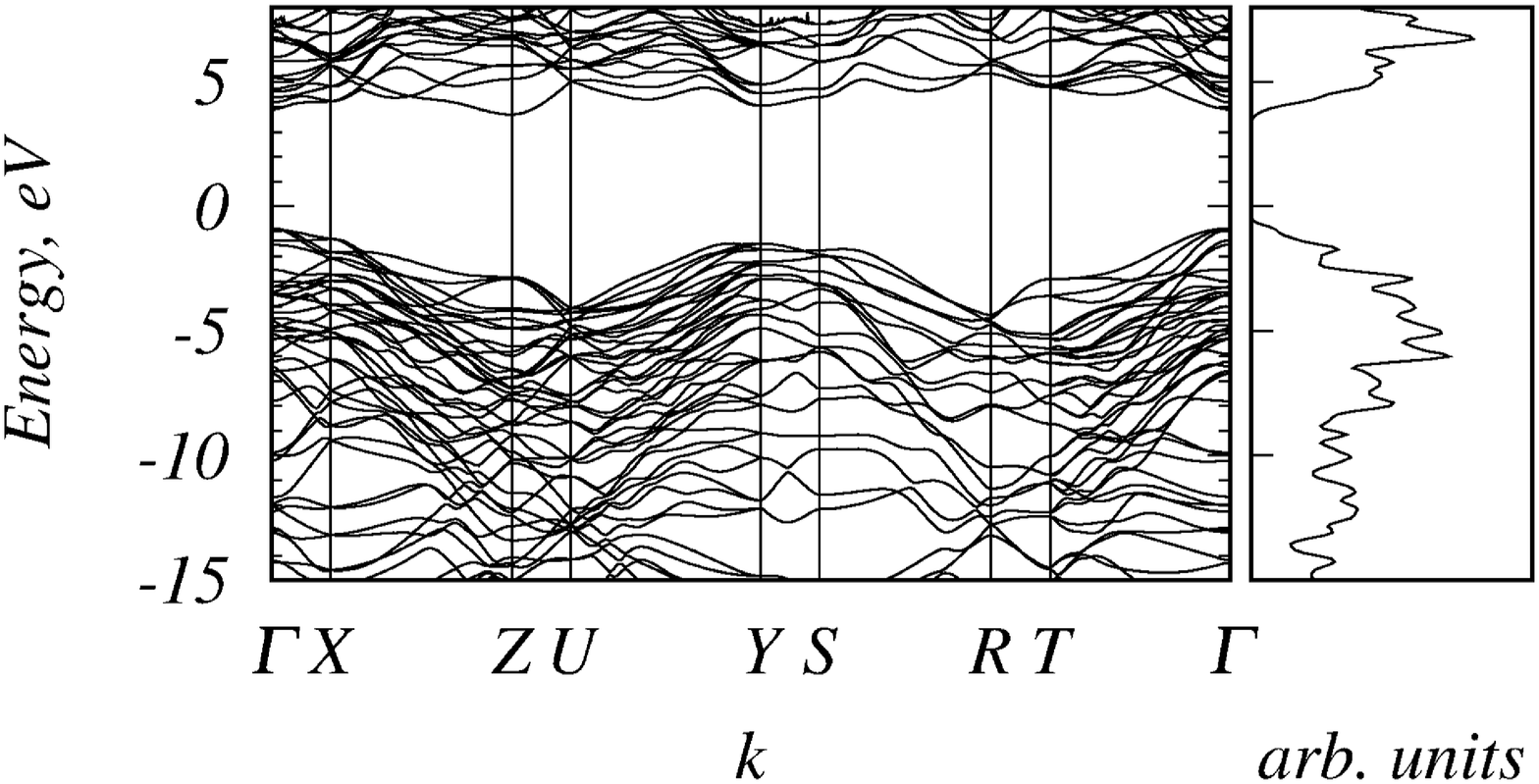}
\includegraphics[width=.3\textwidth, angle=270, clip=]{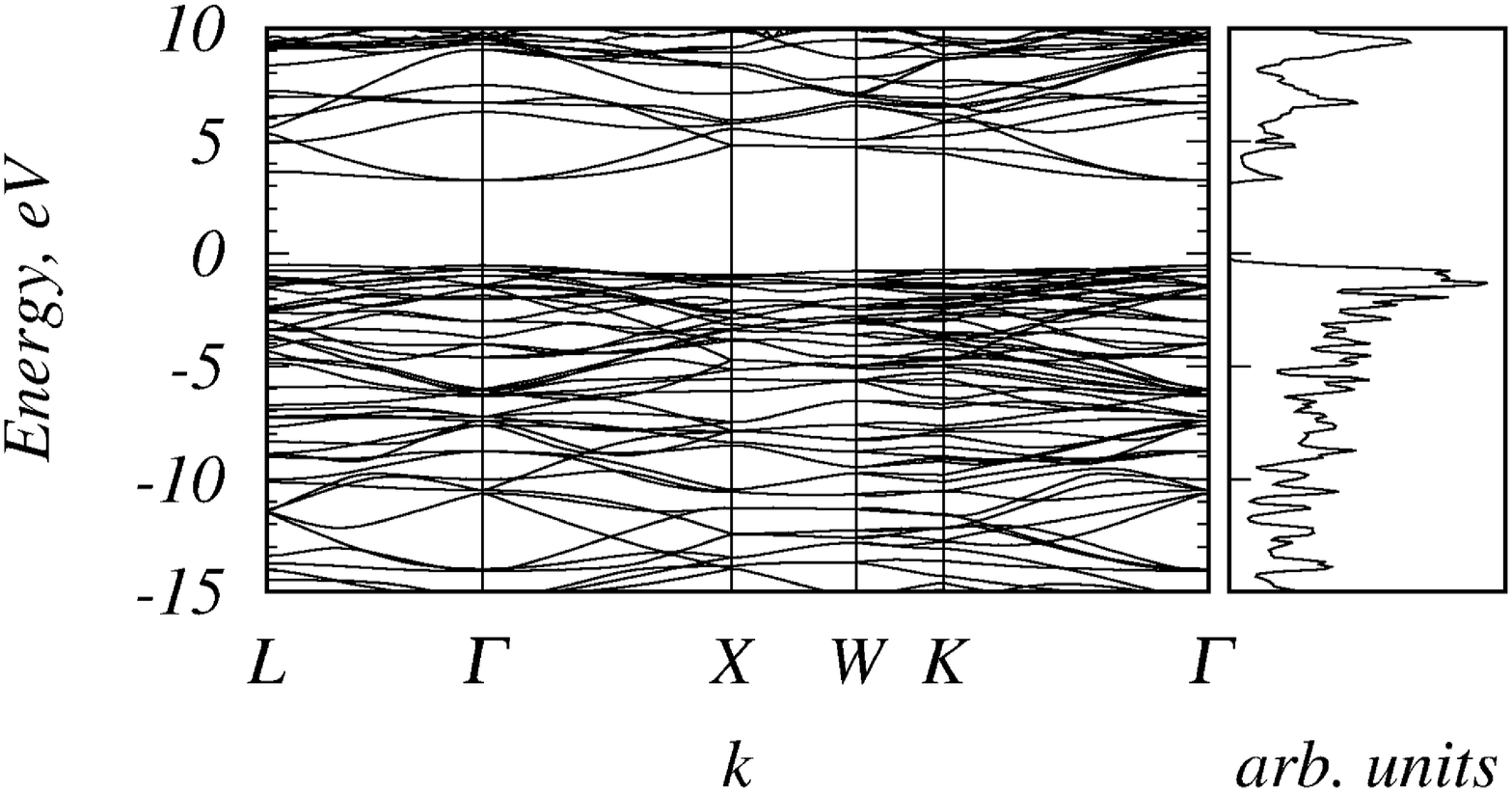}
\includegraphics[width=.3\textwidth, angle=270, clip=]{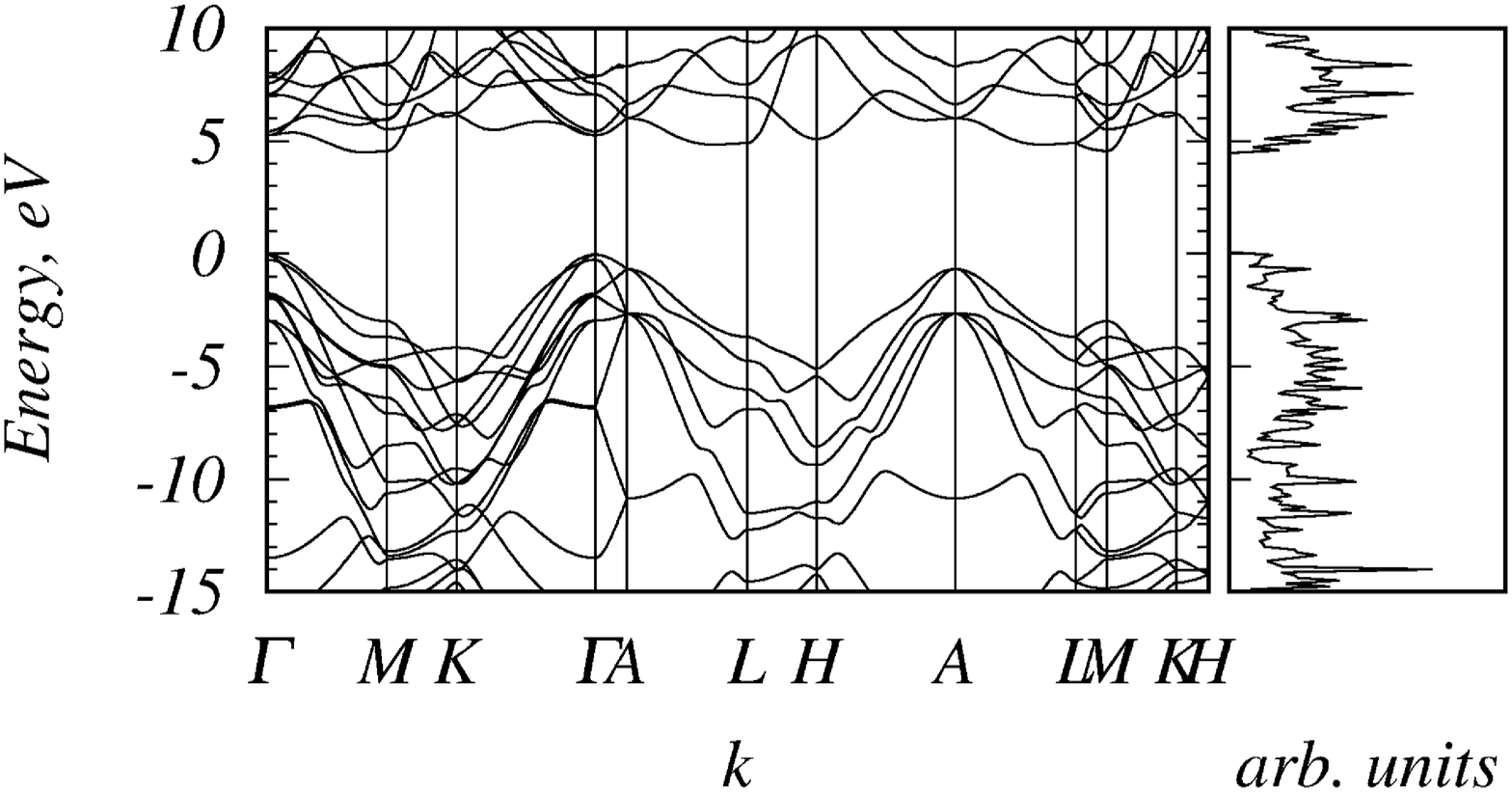}
\caption{Electronic band structures and density of states of
  allotropes. Left column: diamond, SiC12, and mtn. Right column:
lonsdaleite, C28, and 4H-diamond}\label{fig:3}
\end{center}
\end{figure}

Then, we calculated the complex dielectric function for diamond,
lonsdaleite and considered lowest-energy allotropes in the random
phase approximation as it was explained in the Sec. \ref{methods}.
In the Figs.~\ref{fig:2} and~\ref{fig:3} we present our predictions
for diamond obtained with PBE exchange-correlation functional (bold
solid line) and HSE06 functional (dashed line) together with
experimental data (solid line) from the work [1] and PBE based
predictions for lonsdaleite (dash-dotted line). We find a good
agreement between theory and experiment within PBE functional
calculations in the position and highness of the peaks, while with
the HSE06 functional calculations, although providing correct
optical band gaps, leads to the shift of the peaks approximately to
1 eV towards the high frequencies. Also, our results for cubic
diamond and lonsdaleite are consistent with the earlier calculations
in the work [32].

The predictions for allotropes C28, mtn, sic are presented in the
Fig.~\ref{fig:3} in comparison with diamond.The most of new allotropes (SiC12,
H-Carbon, C28) demonstrate the very similar behavior of   to diamond
or lonsdaleite in the optical and UV region, so they do not lose the
desirable optical parameters of diamond. In contrast, the mtn shows
2 times shorter and the smoothest peaks and has the smallest band
gap in comparison to other allotropes, but the position of the peaks
remains the same. Such a way, a presence of mtn in the
polycrystalline sample can contribute a visual opacity.

The effect of the anisotropy of the optical properties for non-cubic
structures is found to be small, so we do not show the dependence of
the dielectric tensor from the crystal directions in the figures,
but only the average values as defined by formula (2.6).

\section{Conclusion}
In summary, the first-principles quantum-mechanical calculations
have been performed to obtain the frequency spectra and optical
properties of lowest-energy $sp^3$ carbon allotropes, such as: cubic
diamond, lonsdaleite, SiC12, 4H-diamond, C28 and mtn. We have
obtained Raman and IR spectra for all discussed allotropes and study
their peculiarities. The electronic structure as well as the linear
photon energy-dependent complex dielectric functions and related
optical properties were computed. Our investigations are beneficial
to the experimental search and to the practical applications of
these hypothetic carbon allotropes in IR, visible and UV photonics
[33,34].

\section{Acknowledgements}
The work was partially funded by the Ministry of Education and
Science of Russia under Competitiveness Enhancement Program of
Samara University for 2013-2020, project 3.5093.2017/8.9.

\begin{table}[p]
\caption{Lattice parameters for carbon allotropes calculated with
POB-TZVP basis set and PBE exchange-correlation
  functional.}

\begin{tabular}{|c|c|c|c|c|c|}
\hline Structure & N & a, {\AA} & b, {\AA} & c, {\AA} & Atomic
positions
 \\  \hline
 diamond & 227 &3.569 & & &(0.125, 0.125, 0.125)\\ \hline
 lonsdaleite & 194& 2.507& & 4.169 &(0.333, 0.667, 0.062) \\ \hline
 mtn & 227 & 9.625 & & &(0.880, 0.067, 0.683)\\
 &&&&& (0.125, 0.125, 0.125)\\
 &&&&& (0.033, 0.033, 0.716)\\ \hline
 SiC12 &166 & 2.511& & 24.822& (0.667, 0.333, 0.740)\\
 &&&&& (0.667, 0.333, 0.572) \\
 &&&&& (0.667, 0.333, 0.511) \\
 &&&&& (0.000, 0.000, 0.656)\\ \hline
 C28 & 55 & 7.298 & 7.728 & 2.535 &(0.328, 0.039, 0.000)\\
 &&&&&(0.376, 0.151, 0.500)\\
 &&&&&(0.207, 0.270, 0.5000)\\
 &&&&&(0.239, 0.384, 0.000)\\
 &&&&&(0.015, 0.196, 0.500)\\
 &&&&&(0.460, 0.408, 0.000) \\ \hline
 4H-diamond & 195 &2.534 & & 8.355& (0.000, 0.000, 0.093)\\
 &&&&&(0.667, 0.333, 0.155) \\ \hline
\end{tabular}\label{Table:6}
\end{table}

\begin{table}[p]
\caption{Lattice parameters for carbon allotropes calculated with
Pople's 6-21 basis set and B3LYP exchange-correlation
   functional.}

\begin{tabular}{|c|c|c|c|c|c|}
\hline Structure & N & a, {\AA} & b, {\AA} & c, {\AA} & Atomic
positions
 \\  \hline
 diamond & 227 &3.594 & & &(0.125, 0.125, 0.125)\\ \hline
 lonsdaleite & 194& 2.527& & 4.205 &(0.333, 0.667, 0.062) \\ \hline
 mtn & 227 & 9.688 & & &(0.880, 0.067, 0.683)\\
 &&&&& (0.125, 0.125, 0.125)\\
 &&&&& (0.033, 0.033, 0.716)\\ \hline
 SiC12 &166 & 2.534& & 25.067& (0.333, 0.667, 0.073)\\
 &&&&& (0.333, 0.667, 0.916) \\
 &&&&& (0.333, 0.667, 0.000) \\
 &&&&& (0.667, 0.333, 0.656)\\ \hline
 C28 & 55 & 7.301 & 7.735 & 2.538 &(0.398, 0.039, 0.000)\\
 &&&&&(0.376, 0.151, 0.500)\\
 &&&&&(0.268, 0.270, 0.5000)\\
 &&&&&(0.239, 0.384, 0.000)\\
 &&&&&(0.015, 0.196, 0.500)\\
 &&&&&(0.460, 0.408, 0.000) \\ \hline
 4H-diamond & 195 &2.513 & & 8.266& (0.000, 0.000, 0.093)\\
 &&&&&(0.667, 0.333, 0.156) \\ \hline
\end{tabular}\label{Table:7}
\end{table}

\begin{table}
\caption{\label{Table:1} Energy difference per atom relatively
diamond, density and bulk modulus of the allotrope (PBE, POB-TZVP).}

\begin{tabular}{|c|c|c|c|}
\hline
Structure & $\triangle E/atom$, eV & $\rho$, g/cm$^3$ & $B$, GPa \\
\hline diamond & 0.00 & 3.51 & 443 \\ \hline 4H-diamond & 0.01 &
3.61 & 445 \\ \hline SiC12 & 0.01 & 3.53 & 444 \\ \hline lonsdaleite
& 0.03 & 3.52 & 445 \\ \hline C28 & 0.07 & 3.35 & 427 \\ \hline mtn
& 0.11 & 3.06 & 383  \\  \hline
\end{tabular}
\end{table}

\begin{table}
\caption{\label{Table:2} Energy difference per atom relatively
diamond, density and bulk modulus of the allotrope (B3LYP, 6-21G).}
\begin{tabular}{|c|c|c|c|}
\hline
Structure & $\triangle E/atom$, eV & $\rho$, g/cm$^3$ & $B$, GPa \\
\hline diamond & 0.00 & 3.43 & 428 \\ \hline 4H-diamond & 0.01 &
3.53 & 429 \\ \hline SiC12 & 0.01 & 3.43 & 428 \\ \hline lonsdalete
& 0.03 & 3.43 & 429 \\ \hline C28 & 0.07 & 3.37 & 413 \\ \hline mtn
& 0.08 & 2.98 & 371 \\ \hline
\end{tabular}
\end{table}

\begin{table}
\caption{\label{Table:3} High-frequency refractive indices along the
different directions.}

\begin{tabular}{|c|c|c|}
\hline
Structure $\backslash$ Basis set & POB-TZVP, PBE& Pople-6-21G, B3LYP  \\
 &$n_{xx}$; $n_{yy}$; $n_{zz}$ & $n_{xx}$; $n_{yy}$; $n_{zz}$ \\ \hline diamond & 2.40; 2.40; 2.40 & 2.34; 2.34; 2.34  \\
\hline 4H-diamond & 2.37; 2.37; 2.43 & 2.32; 2.32; 2.34
\\ \hline SiC12 & 2.38; 2.38; 2.42 & 2.32; 2.32; 2.36  \\ \hline lonsdaleite & 2.36; 2.36; 2.40 &
2.30; 2.30; 2.36 \\ \hline C28 & 2.35; 2.38; 2.43 & 2.29; 2.32; 2.36 \\
\hline mtn & 2.11; 2.11; 2.11 & 2.10; 2.10; 2.10  \\  \hline
\end{tabular}
\end{table}

\begin{table}
\caption{\label{Table:4} IR active modes (Pople-6-21G, B3LYP).}

\begin{tabular}{|c|c|}
\hline
Structure & IR active modes\\
 \hline SiC12 & $E_u$(425), $E_u$(578), $A_{2u}$(728), $A_{2u}$(1075), $E_u$(1271), $A_{2u}$(1330) \\
\hline 4H-diamond & $E_{1u}$(1248), $A_{2u}$(1316)
\\   \hline C28 &  $B_{2u}$(443), $B_{2u}$(539), $B_{3u}$(547), $B_{1u}$(586), $B_{1u}$(719),
$B_{3u}$(742),\\
& $B_{2u}$(786),  $B_{2u}$(862), $B_{3u}$(878), $B_{2u}$(962),  $B_{3u}$(1000),  $B_{2u}$(1002),\\
& $B_{1u}$(1071), $B_{3u}$(1074), $B_{3u}$(1172), $B_{3u}$(1203), $B_{2u}$(1211), $B_{1u}$(1220),\\
& $B_{2u}$(1239), $B_{1u}$(1260), $B_{2u}$(1278), $B_{3u}$(1279), $B_{3u}$(1298), $B_{3u}$(1352),\\
& $B_{2u}$(1356),  $B_{2u}$(1400)\\
\hline mtn & $F_{1u}$(735), $F_{1u}$(871), $F_{1u}$(891),
$F_{1u}$(1059), $F_{1u}$(1144), $F_{1u}$(1223) \\  \hline
\end{tabular}
\end{table}

\begin{table}
\caption{\label{Table:5} Band gaps for carbon allotropes.}

\begin{tabular}{|c|c|c|c|c|}
\hline Structure & PBE, indirect & HSE06, indirect & PBE, direct &
HSE06, indirect
 \\  \hline
 diamond & 4.67 & 5.38 & 5.64 & 7.04 \\ \hline
 lonsdaleite & 3.34 & 4.91& 4.96 & 6.37 \\ \hline
 mtn & 3.76 & 5.09 & 3.76 & 5.03 \\ \hline
 SiC12 & 4.44 & 5.64 & 5.25 & 6.65 \\ \hline
 C28 & 4.77 & 5.96 & 4.77 & 6.06 \\ \hline
 4H-diamond & 4.53 & 5.73 & 5.29 & 6.68 \\ \hline
\end{tabular}
\end{table}

\begin{thebibliography}{99}
 \bibitem{1} Adachi S. Optical constants of crystalline and amorphous
semiconductors.  New York:  Springer Science and Business Media;
1999.

\bibitem{2} Yang N. Novel aspects of diamond (Topics in Applied Physics,
121). Switzerland: Springer International Publishing; 2015.

\bibitem{3} Zaitsev AM. Optical properties of diamond. Berlin:
Springer-Verlag; 2001.

\bibitem{4} Bundy FP, Kasper JS. Hexagonal Diamond—A New Form of Carbon
J Chem Phys 1967; 46: 3437-3446.

\bibitem{5} Frondel C. and Marvin UB. Lonsdaleite, a Hexagonal Polymorph
of Diamond. Nature 1967; 214: 587-589.

\bibitem{6} Hirai H, Kenichi K. Modified phases of diamond formed under
shock compression and rapid quenching. Science 1991; 253(5021):
772-774.

\bibitem{7} Hongliang H, Sekine T, Kobayashi T. Hexagonal diamond
synthesis on h-GaNh-GaN strained films. Appl Phys Lett 2002; 81:
610.

\bibitem{8} Mao WL, Mao HK, Eng PJ, Trainor TP, Newville M, Kao CC,
Heinz DL, Shu J, Meng Y and Hemley RJ. Bonding changes in compressed
superhard graphite. Science 2003; 302: 425-427.

\bibitem{9} Hoffmann R, Kabanov AA, Golov AA, Proserpio DM. Angew Chem
Int Ed 2016; 55: 10962-10977.

\bibitem{10} Hu M, Huang Q et al. Superhard and high-strength
yne-diamond semimetals. Diamond and Related Materials. 2014; 46(0):
15-20.

\bibitem{11} Wang JT, Chen C. Mechanism for direct conversion of
graphite to diamond. Phys Rev B 2011; 84(1): 012102.

\bibitem{12}  Baburin IA, Proserpio DM, Saleev VA, Shipilova AV. From
zeolite nets to $sp^3$ carbon allotropes: A topology-based
multiscale theoretical study. Physical Chemistry Chemical Physics.
2015; 17(2): 1332-1338.

\bibitem{13} Nesper R., Vogel K, et al. Hypothetical Carbon
Modifications Derived from Zeolite Frameworks. Angewandte
Chemie-International Edition in English 1993; 32(5): 701-703.

\bibitem{14} Hohenberg P, Kohn W. Inhomogeneous Electron Gas. Phys Rev
1964; 136: B864-B871.

\bibitem{15} Kohn W, Sham LJ. Self-Consistent Equations Including
Exchange and Correlation Effects.  Phys Rev 1965; 140: A1133-A1138.

\bibitem{16} Dovesi R et al. A program for the ab initio investigation
of crystalline solids. Int J Quantum Chem 2014: 114: 1287-1317.

\bibitem{17} Kresse G, Furthm\"{u}ller J. Efficient iterative schemes for ab
initio total-energy calculations using a plane-wave basis set. Phys
Rev B 1996; 54:11169-11186.

\bibitem{18} Pascale F, Zicovich-Wilson CM, Lopez F, Civalleri B,
Orlando R, Dovesi R. The calculation of the vibration frequencies of
crystalline compounds and its implementation in the CRYSTAL code. J
Comput Chem 2004; 25: 888-897.

\bibitem{19}   Maschio L, Kirtman B, R\'{e}rat M, Orlando R, and Dovesi R.
Ab initio analytical Raman intensities for periodic systems through
a coupled perturbed Hartree-Fock/Kohn-Sham method in an atomic
orbital basis. J Chem Phys 2013; 139: 164101.

\bibitem{20} Ferrero M, Rerat M, Orlando R, Dovesi R. The calculation of
static polarizabilities in 1-3D periodic compounds. The
implementation in the CRYSTAL code. J Comput Chem 2008; 29:
1450-1459.

\bibitem{21}  Perdew JP, Burke K, Ernzerhof M. Generalized Gradient
Approximation Made Simple. Phys Rev Lett 1996; 77: 3865 -3868.

\bibitem{22} Peintinger MF, Oliveira DV, Bredow T. Consistent Gaussian
Basis Sets of Triple-Zeta Valence with Polarization Quality for
Solid-State Calculations. J Comp Chem 2013; 34: 451-459.

\bibitem{23} Becke AD. Density-functional thermochemistry. 3. The role
of exact exchange. J Chem Phys 1993; 98: 5648-5652.

\bibitem{24} Gordon MS, Binkley JS, Pople JA, Pietro WJ, Hehre WJ.
Self-Consistent Molecular Orbital Methods. 22. Small Split-Valence
Basis Sets for Second-Row Elements. J Am Chem Soc 1982; 104:
2797-2803.

\bibitem{25}  Baima J, Zelferino A, Olivero P, Erba A, Dovesi R. Raman
spectroscopic features of the neutral vacancy in diamond from ab
ignition quantum-mechanical calculations. Phys Chem Chem Phys 2016;
18: 1961-1968.

\bibitem{26}  Heyd J, Scuseria GE, Ernzerhof M. Hybrid functionals
on a screened Coulomb potential. J. Chem. Phys. 2006; 77: 219906.

\bibitem{27} Isaenko S, Shumilova T. Thermostimulated Raman spectrum
dynamics of lonsdaleite. Geoph Res Abs 2012; 14:608.

\bibitem{28} Goryainov SV, Likhacheva AY, et al. Raman identification of
lansdaleite in Popigai impactites. J Raman Spectrosc 2014;
45:305-313.

\bibitem{29}  Filik J, Harvey JN, Allan NL, May PW. Raman spectroscopy
of nanocrystalline diamond: An ab initio approach. Phys Rev B 2006;
74:035423.

\bibitem{30} Wu BR, Xu J. Total energy calculations of the lattice
properties of cubic and hexagonal diamond. Phys Rev B 1998; 57:
13355-13359.

\bibitem{31} Denisov VN, Mavrin BN, et. al. First-principles, UV Raman,
X-ray diffraction and TEM study of the structure and lattice
dynamics of the diamond-lonsdaleite system. Diamond and Relared
Matherials 2011; 20:951-953.

\bibitem{32} Wang Z, Zhang RJ, Zheng YX, et al. Electronic and optical
properties of novel carbon allotropes. Carbon 2016; 101: 77-85.

\bibitem{33}  Kaminskii AA, Ral'chenko VG, Yoneda H, Bol'shakov AP,
Inyushkin AV. Stimulated Raman scattering-active isotopically pure
12Ñ and 13Ñ diamond crystals: A milestone in the development of
diamond photonics. JETP Letters  2016; 104(5): 347–352.

\bibitem{34} Salvatori S, GirolamiM, Oliva P,Conte G, Bolshakov
A,Ralchenko V, Konov V. Diamond device architectures for UV laser
monitoring. Laser Phys. 2016; 26: 084005.
\end{thebibliography}
\end{document}